\documentclass[aps,prx,superscriptaddress,twocolumn,amsmath,longbibliography]{revtex4-2}

\usepackage{graphicx, color}
\usepackage{xcolor}
\usepackage{amsfonts,amssymb,amsmath}
\usepackage{siunitx}
\usepackage{bm}
\usepackage{bbold}
\usepackage{braket}
\usepackage{mathtools}
\usepackage{dsfont}
\usepackage{lineno}
\usepackage{bigints}
\usepackage[normalem]{ulem}
\usepackage{enumitem}
\usepackage{amsfonts}

\renewcommand{\vec}[1]{\boldsymbol{#1}}

\renewcommand{\vec}[1]{\boldsymbol{#1}}

\begin{document}

\title{Generic Chiral Anomaly and Planar Hall Effect in a Non-Weyl System}

\author{Yongjian Wang}
\affiliation{Physics Institute II, University of Cologne, Z\"ulpicher Str. 77, 50937 K\"oln, Germany}

\author{Alexander Wowchik}
\affiliation{Institute for Theoretical Physics, University of Cologne, Z\"ulpicher Str. 77, 50937 K\"oln, Germany}

\author{Thomas B\"omerich}
\affiliation{Institute for Theoretical Physics, University of Cologne, Z\"ulpicher Str. 77, 50937 K\"oln, Germany}

\author{A. A. Taskin}
\affiliation{Physics Institute II, University of Cologne, Z\"ulpicher Str. 77, 50937 K\"oln, Germany}

\author{Achim Rosch}
\affiliation{Institute for Theoretical Physics, University of Cologne, Z\"ulpicher Str. 77, 50937 K\"oln, Germany}

\author{Yoichi Ando}
\email[]{ando@ph2.uni-koeln.de}
\affiliation{Physics Institute II, University of Cologne, Z\"ulpicher Str. 77, 50937 K\"oln, Germany}

\begin{abstract}
The condensed-matter version of the chiral anomaly describes how 
electrons are pumped from a Weyl node with negative chirality to a Weyl node with positive chirality using 
parallel electric and magnetic fields. 
%Since Weyl cones have definite chiralities, the chiral anomaly was predicted for Weyl semimetals with signatures being negative longitudinal magnetoresistance (LMR) and planar Hall effect (PHE), both of which have been experimentally observed. 
Key experimental signatures are a  negative longitudinal magnetoresistance (LMR) and the planar Hall effect (PHE), both of which have been experimentally observed. Here, we show that the chiral anomaly explains key features of magnetotransport in the nodal-line semimetal ZrTe$_5$ despite the absence of Weyl points.
The anomaly physics applies generically to materials in the quantum limit, when electron transport becomes quasi-one-dimensional, provided that Fermi velocities remain sufficiently large. This explains not only the negative LMR but also the PHE with a gigantic Hall angle and a highly unusual magnetic-field-angle dependence in ZrTe$_5$.
\end{abstract}
\maketitle

\section{Introduction}

In high-energy physics, the Adler–Bell–Jackiw anomaly describes how quantum corrections break the chiral symmetry for massless Dirac or Weyl particles when these are coupled to electromagnetic fields. In condensed matter physics, an analogous phenomenon occurs in Weyl semimetals: applying parallel electric and magnetic fields transfers charge from a Weyl node with negative chirality to one with positive chirality \cite{Armitage2018}. 
%The chiral anomaly in Weyl semimetals is a condensed-matter version of the Adler-Bell-Jackiew anomaly manifested in the violation of chiral symmetry due to the coupling of chiral particles to electromagnetic fields \cite{Armitage2018}. When parallel electric and magnetic fields are applied to a Weyl semimetal, the Weyl fermions on a pair Weyl cones having opposite chiralities are pumped from one cone to the other, breaking chiral symmetry. 
The experimental signature of this Weyl-fermion pumping is an enhanced current along the electric field $E$, leading to an additional electrical conductivity that is observed as a negative longitudinal magnetoresistance (LMR) \cite{Nielsen1983}. The original theory by Nielsen and Ninomiya considered the Landau quantization due to magnetic fields and assumed that only the lowest Landau level (LLL) is occupied \cite{Nielsen1983}. Later it was shown that one can obtain the same result in a semiclassical theory without considering the Landau quantization \cite{Son2013}. The semiclassical theory also predicts a planar Hall effect (PHE) to accompany the chiral anomaly due to  Berry-curvature effects \cite{Burkov2017, Nandy2017}.

The negative LMR due to the chiral anomaly has been observed not only in Weyl semimetals TaAs, NbP, TaP, WTe$_2$, and Co$_3$Sn$_2$S$_2$ \cite{Huang2015, Zhang2016, Wang2016, Arnold2016, Wang2016a, Liu2018}, but also in three-dimensional (3D) Dirac semimetals Na$_3$Bi, GdPtBi, and Cd$_3$As$_2$ \cite{Xiong2015, Hirschberger2016, Li2015, Li2016a, Ong2021}, where the magnetic field splits the 3D Dirac cone into a pair of Weyl cones \cite{Armitage2018}  making the system a Weyl semimetal in magnetic fields. A similar negative LMR has been observed in ZrTe$_5$ and interpreted to be also due to the chiral anomaly \cite{Li2016}, but the case in ZrTe$_5$ is not as simple, because the Dirac cone of the effective 3D Dirac Hamiltonian of ZrTe$_5$ \cite{Weng2014, RYChen2015, Wang2022, Wang2023, Wang2024} does {\it not} split into Weyl cones in the magnetic field $B$ along the $a$-axis \cite{RYChen2015}, along which the electric field $E$ was applied. A similar negative LMR without Weyl cones has been reported for PdCoO$_2$ \cite{Kikugawa2016} and SrAs$_3$ \cite{Li2018b}, and it was proposed that the formation of quasi-one-dimensional (quasi-1D) transport channels due to Landau quantization in the quantum limit (and the resulting isolation of the Fermi momenta) is the origin \cite{Kikugawa2016, Li2018b}. 
%It is worth noting that in both ZrTe$_5$ and PdCoO$_2$, the quantum limit is achieved in experiments. 

In the present paper, we extend the idea of Refs. \cite{Kikugawa2016, Li2018b} and theoretically show that the origin of the negative LMR in the quantum limit of a non-Weyl system is actually the same as the chiral anomaly in Weyl semimetals, even down to the universal prefactor of the chiral-magnetic term. This is essentially because the same physics stemming from quasi-1D transport in the LLL lies behind the negative LMR, and hence this regime can be generically called ``chiral-anomaly regime" in many materials. We observed negative LMR in the nodal-line semimetal ZrTe$_5$ for transport along both $a$ and $b$ axes in the absence of Weyl nodes, and our theoretical calculations based on this generic chiral-anomaly scenario within the self-consistent Born approximation indeed reproduce the experimentally-observed features of LMR. We further observed a PHE in ZrTe$_5$ with a gigantic Hall angle and an unusual magnetic-field-direction dependence when $B$ is rotated in the $ab$ plane, which is in contrast to the usual PHE which presents $\sin(2\theta)$ dependence on the angle $\theta$ between $E$ and $B$ \cite{Kokado2012, Taskin2017, Burkov2017, Nandy2017, Li2018, Kumar2018, Li2018a, Wu2018, Chen2018, Liang2019, Li2019, Yang2020}. We show that this unconventional PHE is another manifestation of the quasi-1D transport in the chiral-anomaly regime realized in a highly anisotropic medium. This means that, similar to the semiclassical regime of a Weyl semimetal \cite{Burkov2017, Nandy2017}, the chiral anomaly in a non-Weyl system is also accompanied by a PHE, but with a different mechanism.

\section{Chiral anomaly from the LLL}

\subsection{Longitudinal conductivity of a Weyl system}\label{main:chiralAnomaly}

We start with a brief presentation of the chiral anomaly in Weyl semimetals in magnetic fields, which was originally discussed by Nielsen and Ninomiya \cite{Nielsen1983}. For  weak scattering from one Weyl point to another by disorder (or interactions), the chiral anomaly describes that in the presence of both electric and magnetic fields, charge is pumped from one Weyl point to the next \cite{Nielsen1983}
\begin{align}
    \partial_t \rho^+-\partial_t \rho^- =2 \frac{e^3}{4 \pi^2 \hbar^2} \vec E \cdot \vec B-\frac{\rho^+-\rho^-}{\tau}\,,
    \label{main:anomaly}
\end{align}
where $\rho^\pm$ is the charge density at  two Weyl points with opposite chirality, $\pm 1$. Remarkably, the prefactor of the $\vec E \cdot \vec B$ term, the chiral anomaly, is completely universal and independent of all microscopic parameters. The last term describes phenomenologically the scattering of electrons from one Weyl point to the other.

Deep in the quantum limit, i.e. for a large magnetic field, this formula has a simple interpretation: For each Weyl point one obtains per flux quantum a single chiral channel with dispersion $\pm v_F \hbar k$.  Using Newton's law, $\dot p=e E$, and momentum quantization $\Delta p=\hbar \frac{2 \pi}{L}$ in a 1D system of length $L$, one obtains $\partial_t Q^{\pm}=V \partial_t \rho^\pm=\pm e N_{\rm LL} \frac{\dot p}{\Delta p}$, where $V=A L$ is the volume of the system and $N_{\rm LL}=\frac{A}{2 \pi l_B^2}$ the degeneracy of the LL. Thus, one obtains for parallel magnetic and electric fields $\partial_t \rho^\pm=\pm\frac{e^3}{4 \pi^2\hbar^2} E B$, or, equivalently, Eq.~\eqref{main:anomaly}, after adding inter\-nodal scattering.

For a Weyl system in the quantum limit, the electrons at the two Weyl points move with  the Fermi velocity $\pm v_F$ parallel to the magnetic field. Thus, in the stationary state one obtains the current density 
\begin{align}
j = v_F (\rho^+-\rho^-)=\frac{ \tau v_F e^3}{2 \pi^2 \hbar^2} E B,
\end{align}
or, equivalently, for the longitudinal conductivity
\begin{align}
\sigma_\| = \frac{e^2 v_F}{2 \pi^2  l_B^2 \hbar} \tau.
\label{main:sigmapara1}
\end{align}

\subsection{Longitudinal conductivity of a non-Weyl system}
\label{theo1} 

\begin{figure}[t]
	\centering
	\includegraphics[width=8.5cm]{ 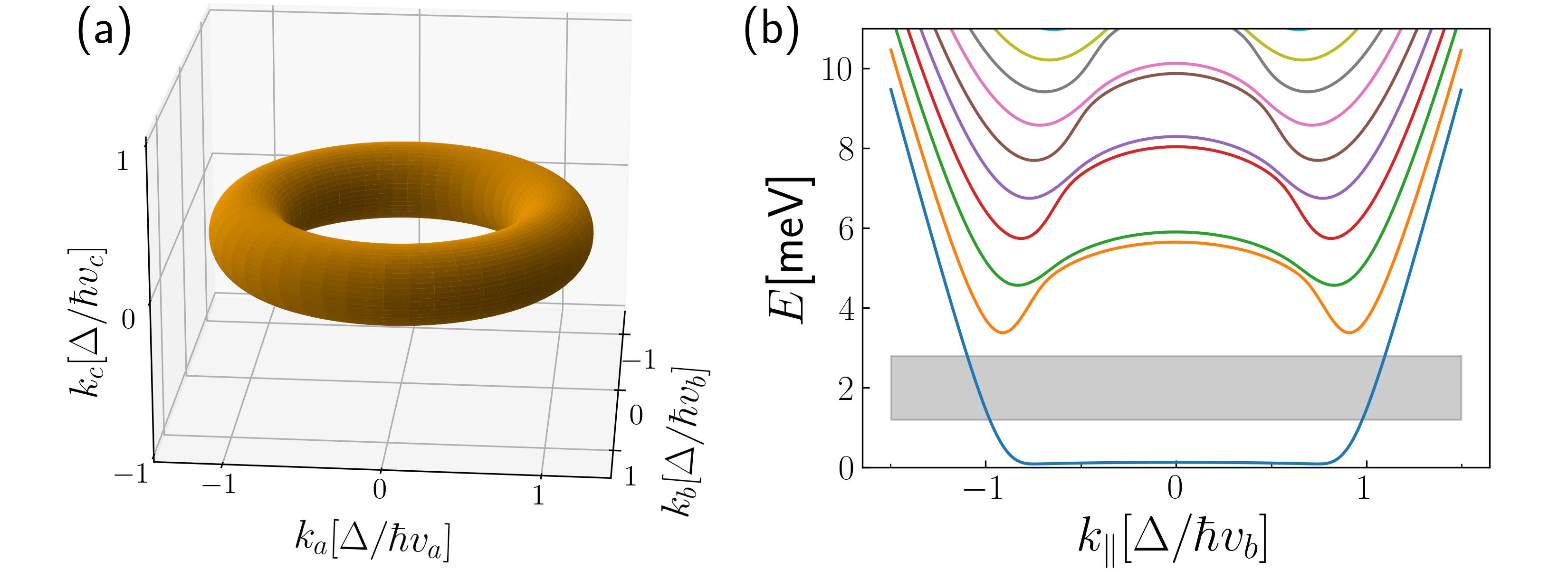}
	\caption{\label{fig:theory1} (a) A slightly doped nodal-line semimetal has a torus shaped Fermi surface. Such a state can be realized in ZrTe$_5$ \cite{Wang2022}, where the nodal line is an ellipse in the $ab$ plane of width $\Delta/v_a$ and $\Delta/v_b$, see \cite{SM}.
    %see App.~\ref{app:th:model}.  
    (b) Landau level energy as function of the momentum parallel to a magentic field  in the $b$ direction ($B_b=0.2\,$T). The lowest Landau level is ultraflat \cite{Wang2023} for momenta smaller than the width of the nodal line,   
    $k_\| \ll \Delta/v_b$. For Fermi energies in the gray-shaded area, transport is well-described by a chiral anomaly.
    The model is particle-hole symmetric and only positive energies are shown for simplicity (the experimental system is hole-doped).
    }
    \end{figure}

Now we develop an analytical theory of longitudinal transport in a non-Weyl system with a low carrier density in the quantum limit. 
As a specific example, we consider a slightly-doped nodel-line semimetal, which has a torus-shaped Fermi surface as sketched in Fig.~\ref{fig:theory1}(a). As has been shown in Ref.~\cite{Wang2022} both by a symmetry analysis and quantum oscillation experiments, such a state can be realized in ZrTe$_5$.

Upon increasing $B$, LLs form and the chemical potential $\mu$ becomes smaller and smaller as the degeneracy within a LL becomes larger. 
When $\mu$ crosses only the  lowest LL, %(which we assume to disperse almost linearly near $k_F$) 
and the scattering rate $\Gamma$ is much smaller than the LL spacing, a very simple theoretical description emerges: per flux quantum of the magnetic field, only a single 1D channel contributes to transport.
This situation is realized for chemical potentials in the grey-shaded area shown in Fig.~\ref{fig:theory1}(b).
Expanding around the two Fermi points with $k=\pm k_F$, the Green function is given by
\begin{align}\label{Geff}
G_k(\omega)\approx \sum_{\sigma=\pm}\frac{|\sigma k_F\rangle \langle \sigma k_F|}{\hbar \omega-\sigma v_F\hbar(k-\sigma k_F)+ i \Gamma_{k_F}}, 
\end{align}
where $k$ is the momentum parallel to the magnetic field and $|\pm k_F\rangle$ encodes the wave function of the LLs at $k=\pm k_F$. 
Within the Born approximation, the imaginary part of the self-energy, projected onto the lowest band is given by 
\small
\begin{align}\label{eq:gammaTot}
    \Gamma_{k_F}&\approx \Gamma^+_{k_F}+\Gamma^-_{k_F}=\frac{\gamma}{v_F \hbar 4 \pi l_B^2} (\alpha^{+}+\alpha^{-}),
\end{align}
\normalsize
where $\Gamma^{\pm}$ describes scattering from $k_F$ to $\pm k_F$, respectively, $\gamma$ parametrizes the amplitude of the disorder potentials, and $\alpha^{\pm}$ are numerical factors arising from the overlap of wavefunctions at the two Fermi points and are obtained by diagonalizing the Hamiltonian of the system, see \cite{SM} for details and definitions. 
%Note that we discuss the regime where the broadening of spectral functions by $\Gamma_{k_F}$ is neglected.  The numerical factors $\alpha^{\pm}$ are properties of the LL wave function and are obtained by diagonalizing the Hamiltonian of the system.

Ignoring vertex correction and using the Kubo formula for a single band, the longitudinal conductivity at $T=0$ in this regime is computed analytically as
\begin{align}
    \sigma_{\parallel}\approx 2 \frac{e^2 \hbar v_F^2}{2 \pi l_B^2} \frac{1}{\pi} \int \frac{dk}{2 \pi} \left(\text{Im}\, g_k(\omega=0)\right)^2\approx \frac{e^2 v_F}{4 \pi^2 l_B^2 \Gamma_{k_F}},
    \label{main:sigmapara2}
\end{align}
where $g_k(\omega)=\frac{1}{\hbar \omega-v_F\hbar k+ i \Gamma_{k_F}}$. 

Identifying $\frac{\hbar}{\tau}=2 \Gamma_{k_F}$, this formula corresponds exactly to Eq.~\eqref{main:sigmapara1}. 
This is not surprising, as the underlying physics is identical \cite{Kikugawa2016}: per flux quantum a 1D channel forms parallel to the magnetic field.
Using Eq.~\eqref{eq:gammaTot} and Eq.~\eqref{main:sigmapara2},
the longitudinal conductivity is given by $\sigma_\| \approx \frac{e^2}{2 \pi \hbar} \frac{v_F^2 \hbar^2}{ \gamma (\alpha^+ +\alpha^-)/2}$, or, if vertex corrections are taken into account \cite{SM} by
\begin{align}\label{eq:condVertex}
      \sigma_{\|}\approx \frac{e^2}{2 \pi \hbar} \frac{v_F^2 \hbar^2}{ \gamma \alpha^-}.
\end{align}

While the anomaly physics is identical in the field range for which Eq.~\eqref{Geff} is valid, for non-Weyl systems it can typically be applied only in an intermediate field range. 
For small magnetic fields, quantities like $\rho^{\pm}$ are not well-defined as long as there are no well-separated Fermi points. 
For very large fields, in contrast,
there is the necessary condition that the broadening by disorder in energy is smaller than the Landau-level spacing $\Delta_\text{LL}$ and the broadening in momentum space is smaller than the distance $\Delta k$ of the points $\pm k_F$
\begin{align}\label{eq:validity}
   \Delta_\text{LL} \gg \Gamma_{k_F}, \qquad   \Delta k \gg \frac{\Gamma_{k_F}}{v_F}.
\end{align}
As we will discuss below, this condition is met for a chemical potential in the gray-shaded area of Fig.~\ref{fig:theory1}(b) but breaks down in the flat-band regime.

%A necessary condition for the validity of this chiral-anomaly picture is that the scattering is sufficiently weak. Only in this case, quantities like $\rho^{\pm}$ are well defined.
%In both the Weyl and non-Weyl systems, this leads to the condition that the distance of the Fermi points in the $k$-space has to be much larger than $1/\ell$, where $\ell$ is the mean-free path from scattering, $\Delta k \gg 1/\ell$, or, equivalently, as $\ell=v_F \tau$,
%\begin{align}
 %   v_F \Delta k \gg \frac{1}{\tau}.
%\end{align}
%In the discussion above, we assumed $\Delta k=2 k_F$ which is of order $\Delta/(v_F\hbar)$ in the chiral anomaly regime ($\Delta$ is the LL spacing), thus we demand $\Delta \gg \frac{\hbar}{\tau}$. Indeed, this is the assumption we made for the chiral-anomaly. 

\section{Experiments}

\begin{figure}[t]
	\centering
	\includegraphics[width=8.5cm]{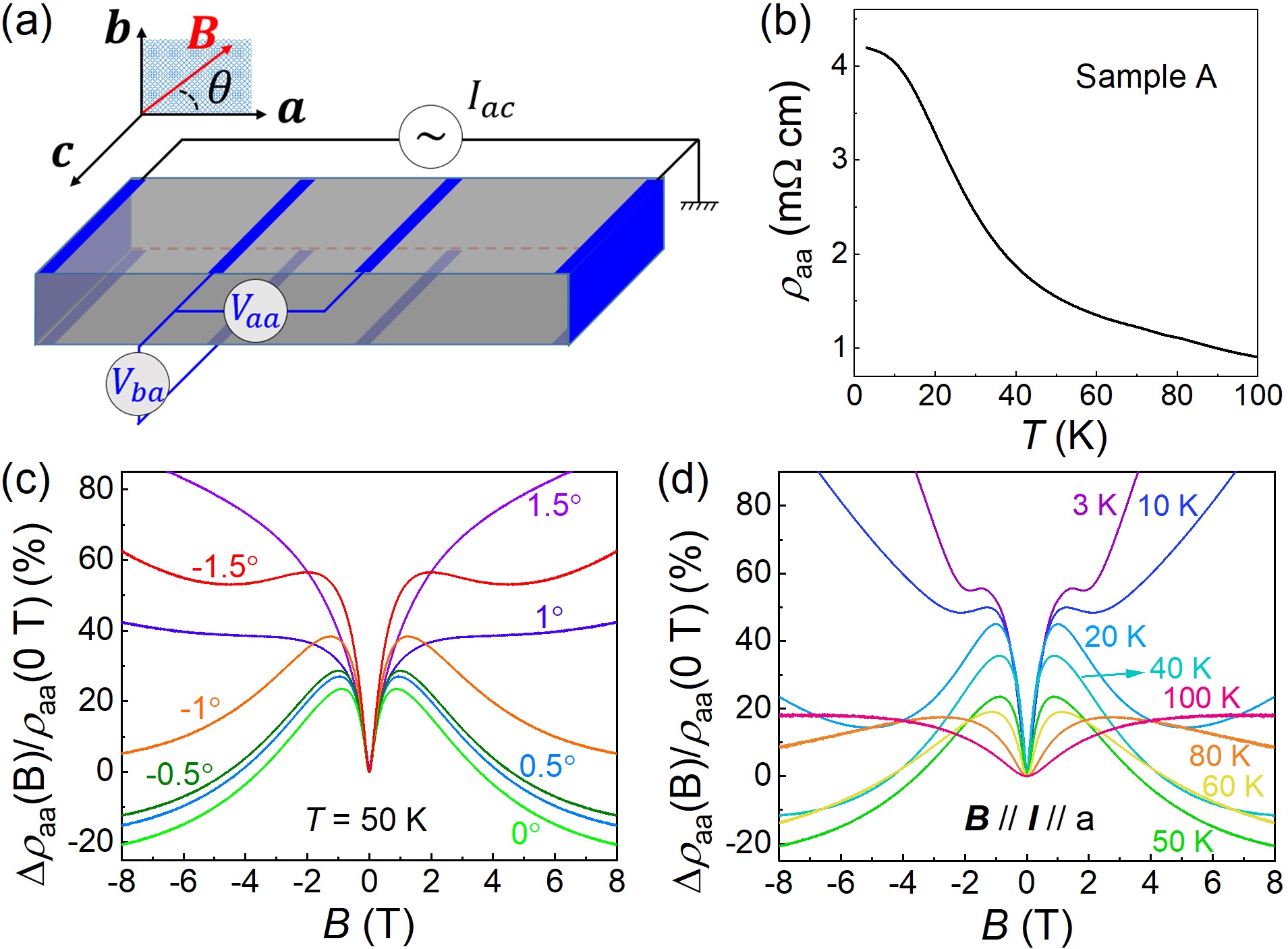}
	\caption{(a) Schematic of the measurement configuration for Sample A. $\rho_{ba}$ was measured with the electrodes made on the top and bottom surfaces of the ZrTe$_5$ crystal. $\theta$ is defined as the angle between $\boldsymbol {B}$ and the $a$-axis in the $ab$ plane, which is the rotation plane. The current was applied along the $a$-axis. (b) Temperature dependence of $\rho_{aa}$ in 0 T. (c) Magnetoresistance measured at 50 K with varying $\theta$ near the $a$-axis. (d) Longitudinal magnetoresistance for $\boldsymbol {B} \parallel \boldsymbol {I} \parallel a$-axis ($\theta$ = 0$^\circ$) measured at various temperatures.} \label{fig:exp1}
\end{figure}

The topological semimetal ZrTe$_5$ studied in this work is a layered van-der-Waals material and each layer consists of ZrTe$_3$ chains that extend along the $a$ axis and are connected via Te atoms in the $c$-direction; hence, the layer forms the crystallographic $ac$ plane that are stacked along the $b$ axis. It is the convention for ZrTe$_5$ to take $a$, $c$, and $b$ axes for the $x$, $y$, and $z$ directions, respectively \cite{Weng2014}. The temperature dependence of the $a$-axis resistivity $\rho_{aa}$ always shows a pronounced peak, and the peak temperature $T_p$ is dependent on how the crystals were grown. It has been elucidated that $T_p$ corresponds to the temperature where the chemical potential (which shifts with temperature) is located at the Dirac point \cite{Xu2018}. While the initial report of the negative LMR was made for a $T_p \approx$ 60 K sample \cite{Li2016}, it was also observed in samples with $T_p \approx$ 0 K \cite{Wang2022, Wang2023}. Since the $T_p \approx$ 0 K samples have very low carrier density of the order of 10$^{16}$--10$^{17}$ cm$^{-3}$ at low temperature \cite{Wang2022} and hence can be easily brought into the quantum limit, we study $T_p \approx$ 0 K samples here [see Fig.~\ref{fig:exp1}(b)]. 
Our previous studies showed that in those $T_p \approx$ 0 K samples inversion symmetry is broken, which splits the 3D Dirac point into a line of nodes and the system becomes a nodal-line semimetal \cite{Wang2022}.

To study the origin of the negative LMR in ZrTe$_5$, we performed magnetotransport measurements while rotating $B$ in the $ab$ plane [see Fig.~\ref{fig:exp1}(a)]. 
This rotation plane is chosen because the 3D Dirac Hamiltonian of ZrTe$_5$ dictates that the nodal-line created by the broken inversion symmetry does not break up into Weyl points in those magnetic fields. We can therefore study the effects of strong magnetic fields in the absence of Weyl cones for $\boldsymbol{B} \parallel ab$-plane. We applied $E$ along the $a$-axis and measured the voltages $V_{aa}$ and $V_{ba}$ that appeared along $a$ and $b$ directions, respectively, in response to the current $\boldsymbol{I} \parallel a$-axis. 

%We observed an unusual PHE in $\rho_{ba}$ whose magnetic-field-angle-dependence deviates strongly from the conventional $\sin(2\theta)$ dependence. We show that this unconventional PHE can be understood with a combination of the strong Fermi-surface anisotropy and the magnetic-field-induced resistivity anisotropy. This implies that the peculiar magnetotransport properties of ZrTe$_5$, including the negative LMR, is caused by the quasi-1D transport channels formed in the quantum limit in a very anisotropic environment. To support this conclusion, our theoretical calculations of LMR based on the self-consistent Born approximation, considering the Landau quantization but without considering the chiral anomaly, reproduce essential features of the experimentally-observed LMR. Besides establishing the understanding of the LMR in ZrTe$_5$, the unconventional PHE discovered here is an interesting phenomenon of its own that occurs in the quantum limit in a strongly anisotropic system.

%\subsection{$\rho_{aa}$}

For sample A (50-$\mu$m wide and 52-$\mu$m thick) used for measuring $\rho_{aa}$ and $\rho_{ba}$, voltage contacts were made on the top and bottom of the naturally cleaved $ac$ surfaces, while current contacts were made to cover the $bc$ surfaces, as depicted in Fig.~\ref{fig:exp1}(a) (see \cite{SM} for details). 
Figures 1(c) and 1(d) show that all the characteristic features of the negative LMR known for ZrTe$_5$ \cite{Li2016, Liang2018} are reproduced in $\rho_{aa}$ of our sample: (i) Positive LMR at very low $B$ which turns into negative at higher $B$; (ii) onset of another positive component at even higher $B$; and (iii) observability of the negative component for a relatively wide temperature range (up to $\sim$100 K) but only in a very limited range of $\theta$ near the $a$-axis (within $\pm$1.5$^{\circ}$). Note that the largest negative LMR was observed at around 50 K, which is different from the $T_p \approx$ 60 K samples \cite{Li2016}.
 
 \begin{figure}[b]
	\centering
	\includegraphics[width=8.5cm]{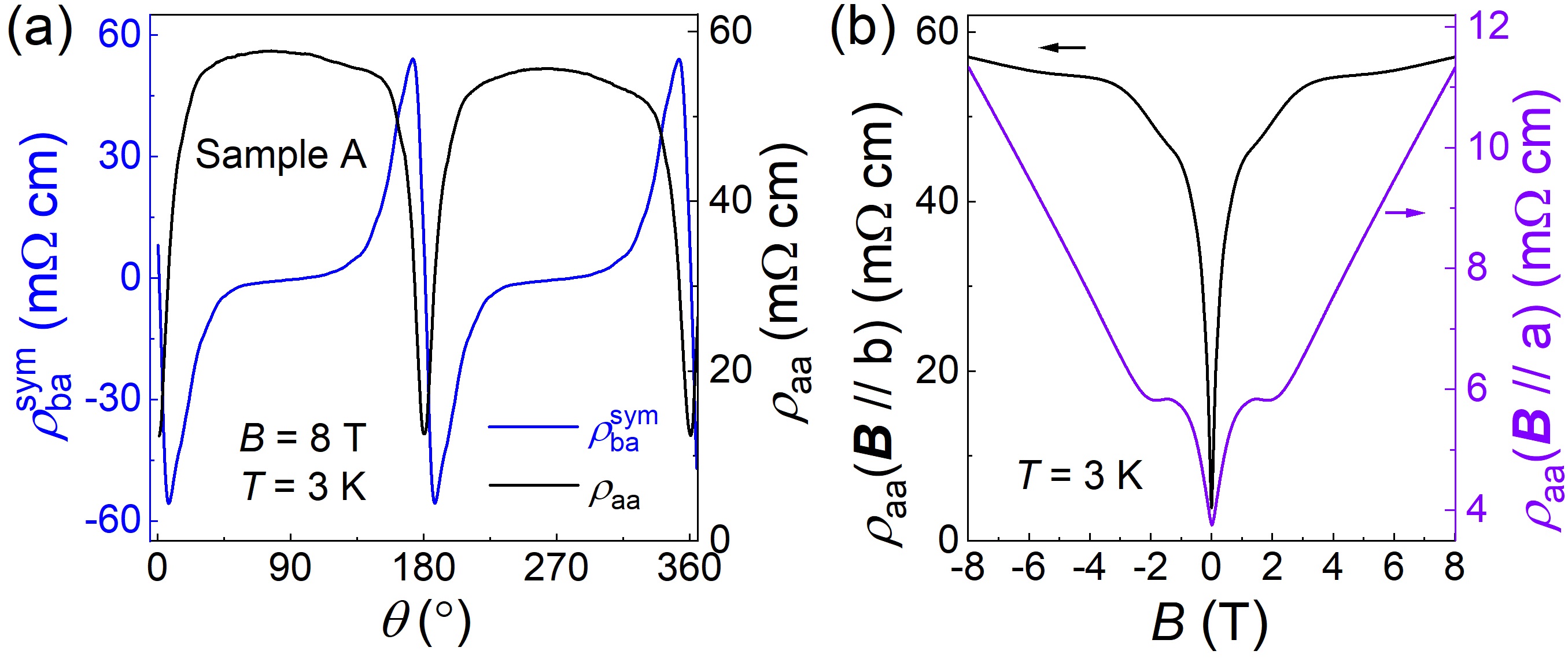}
	\caption{(a) Angular dependencies of $\rho_{aa}$ in 8 T and the $B$-symmetric component of $\rho_{ba}$ in 8 T, measured at 3 K. (b) Behaviors of $\rho_{aa}(B)$ for $\boldsymbol {B} \parallel b$-axis (transverse) and for $\boldsymbol {B} \parallel a$-axis (longitudinal) at 3 K.} \label{fig:exp2}
\end{figure}

Interestingly, an unusual PHE was observed when $B$ was rotated in the $ab$ plane. The observed $\rho_{ba}(B)$ was essentially symmetric in $B$ (see \cite{SM}), which is consistent with the PHE, but its angular dependence shown in Fig.~\ref{fig:exp2}(a) is very unusual, deviating significantly from the conventional $\sin(2\theta)$ behavior \cite{Kokado2012, Taskin2017, Burkov2017, Nandy2017, Li2018, Kumar2018, Li2018a, Wu2018, Chen2018, Liang2019, Li2019, Yang2020}. 
%Note that the PHE due to the chiral anomaly is conventional in terms of its $\theta$-dependence \cite{Burkov2017, Nandy2017}, so the observed $\theta$-dependence for $B \parallel ab$ corroborates the non-chiral-anomaly origin of the PHE. 
The magnitude of the PHE signal in 8 T at 3 K is up to $\sim$54 m$\Omega$cm [Fig.~\ref{fig:exp2}(a)], which is very large in comparison to $\rho_{aa}$ in 0~T ($\sim$4 m$\Omega$cm) and is comparable to the maximum $\rho_{aa}$ shown in the same panel. The latter occurs for $\boldsymbol {B} \parallel b$ and reflects a large positive transverse magnetoresistance (TMR). The peak in the PHE signal in Fig.~\ref{fig:exp2}(a) gives the Hall angle of 86$^{\circ}$, an unusually large value signifying the gigantic nature of this PHE. 

%Although this PHE shows a period of $\pi$, its dependence on $\theta$ is very different from the conventional $\sin(2\beta)$ behavior \cite{Tang2003, Kokado2012, Taskin2017}; note that the PHE due to the chiral anomaly is conventional in terms of its $\theta$-dependence \cite{Burkov2017, Nandy2017}.

%$B$-symmetric component was observed in $\rho_{ba}$ at 3 K in 8 T, and its $\theta$-dependence is plotted in Fig.~\ref{fig:exp2}(a). This component signifies the PHE, because it is observed in the plane spanned by $I$ and $B$, and it is symmetric with $B$. Its magnitude (up to $\sim$54 m$\Omega$cm) is large in comparison to $\rho_{aa}$ in 0 T ($\sim$4 m$\Omega$cm) and is comparable to the maximum $\rho_{aa}$ in the same panel. The latter occurs for $B \parallel b$ and reflects a large positive transverse magnetoresistance (MR). Although this PHE shows a period of $\pi$, its dependence on $\theta$ is very different from the conventional $\sin(2\beta)$ behavior \cite{Tang2003, Kokado2012, Taskin2017}; note that the PHE due to the chiral anomaly is conventional in terms of its $\theta$-dependence \cite{Burkov2017, Nandy2017}.

%\subsection{$\rho_{ba}$}

The observed PHE in the $ab$-plane presents a sharp peak-and-dip behavior when $B$ crosses the $a$-axis (i.e. near $\theta$ = 0$^\circ$ and 180$^\circ$), and the $\theta$-dependence of $\rho_{aa}$ in 8 T shows a correspondingly sharp minima at $\theta$ = 0$^\circ$ and 180$^\circ$. The sharp peak-and-dip structure becomes most pronounced at around 50 K. As a function of $B$, it simply weakens as $B$ is decreased (see \cite{SM}). 

Figure~\ref{fig:exp2}(b) shows the $B$-dependences of $\rho_{aa}$ for $\boldsymbol {B} \parallel b$ (TMR) and for $\boldsymbol {B} \parallel a$ (LMR) plotted together. One can see that the TMR is singular at low fields and saturates above $\sim$3 T, where the quantum limit is reached \cite{Wang2023}. On the other hand, the LMR shows a wiggle at around $\pm$2 T, before the resistivity increases strongly in the flat-band regime. This wiggle is the signature of the negative LMR, which becomes prominent at around 50 K in this $T_p$ = 0~K sample, as one can see in Fig.~\ref{fig:exp1}(d).

\section{Theory of parallel transport}

\begin{figure}[t]
	\centering
	\includegraphics[width=8.5cm]{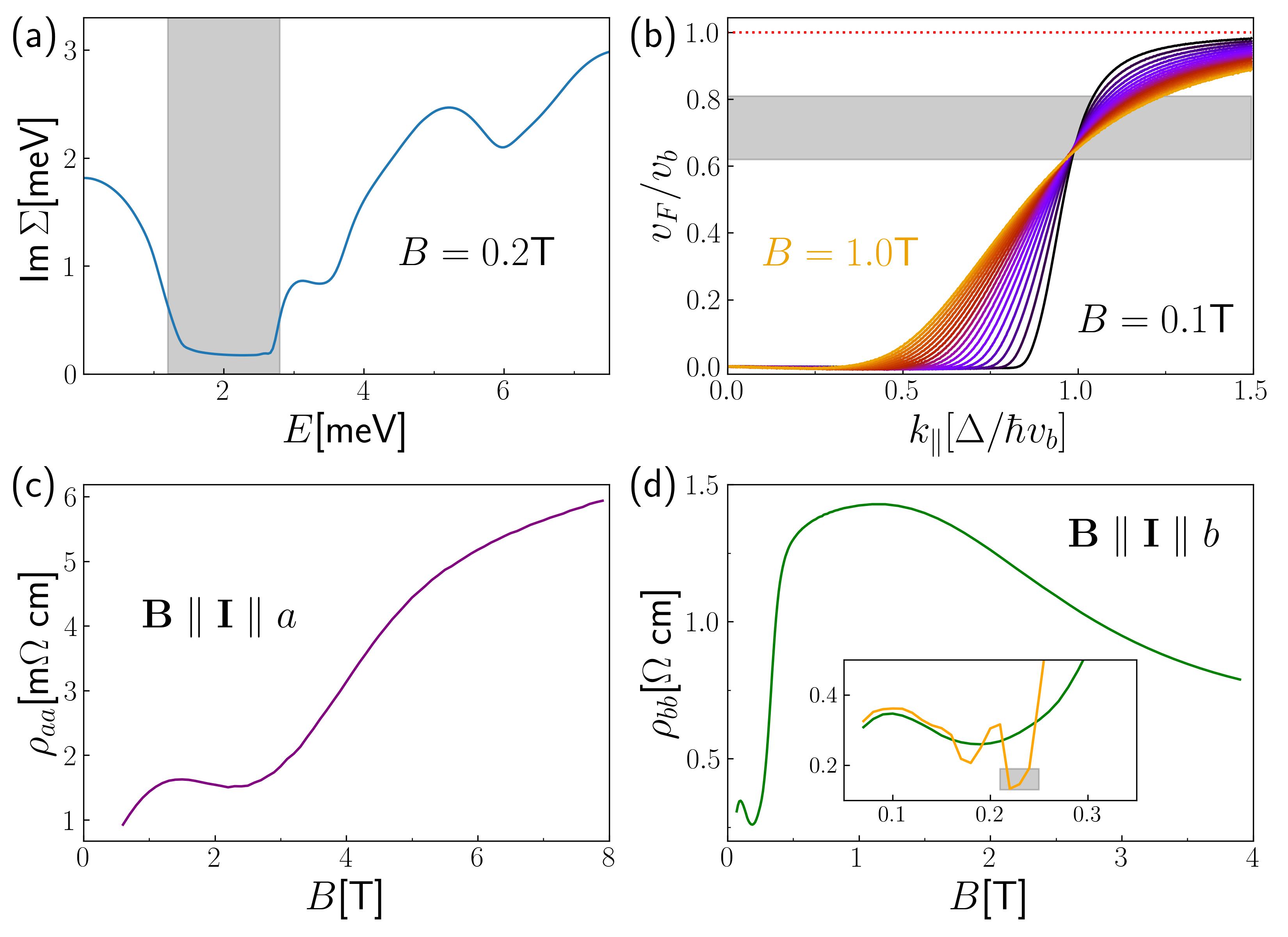}
	\caption{Theory of longitudinal transport. (a) Average scattering rate within self-consistent Born approximation, calculated for $B_b=0.2\,$T, i.e., for the band-structure shown in Fig.~\ref{fig:theory1}(b) ($\gamma=0.06$\,nm$^3$eV$^2$). In the chiral-anomaly regime (shaded region) the scattering rate is more than an order of magnitude smaller than the distance of LLs.
    (b) Velocity of electrons in the lowest Landau level as function of $k_\|$ for magnetic fields in the $b$-direction with  0.1~T$\,\le B_b\le 1\,$T. The velocity is approximately constant in the chiral-anomaly regime.
    (c, d) Calculated longitudinal resistivity $\rho_{aa}$ and $\rho_{bb}$ for $\boldsymbol {B} \parallel a$ and $\boldsymbol {B} \parallel b$, respectively. The curves qualitatively reproduce the negative LMR visible as a dip in $\rho$ for small magnetic fields, see Fig.~\ref{fig:exp2}(b) and Fig.~\ref{fig:exp3}(a), but there are also discrepancies (e.g., an extra downturn in $\rho_{bb}$ for the smallest accessible $B$ fields and a downturn for large $B$), see text.
    The curves are obtained by averaging over density fluctuations, see \cite{SM}. Non-averaged curves (orange line in the inset of panel (d) obtained for $n= 1.1\times 10^{16}\,\text{cm}^{-3}$) show pronounced quantum oscillations when $\mu$ crosses the band minima shown in panel~(a). 
    %The actual $T_p \approx$ 0 K samples of ZrTe$_5$ are slightly hole-doped and the picture in panel (a) should be understood to be inverted for explanation purpose.
} \label{fig:Theory2}
\end{figure}

To understand the origin of the complicated behavior of LMR including the limited appearance of negative LMR, we have performed transport calculations taking into account  the formation of Landau levels.
%based on the bandstructure of ZrTe$_5$. 
The $T_p \approx$ 0 K samples of ZrTe$_5$ are a slightly hole-doped nodal-line semimetal \cite{Wang2022} with a torus-shaped Fermi surface at zero magnetic field as sketched in  Fig.~\ref{fig:theory1}(a). 
As discussed in Ref.~\cite{Wang2023}, the special topological nature of the band structure leads to the formation of an ultraflat band in the presence of a large external magnetic field in either the $a$ or $b$ direction, see Fig.~\ref{fig:theory1}(b).

To develop a quantitative transport theory, we have to take into account two types of disorder effects. First, we model short-ranged scattering by a $\delta$-correlated impurity potential. Second, due to the extremely low electron density, a few ppm per unit cell, screening is inefficient and charged impurities lead to large scale inhomogeneities, which have been argued to be responsible for giant non-linear transport effects observed in Refs.~\cite{Wang2022,Wang2023}.

The minimal theory able to compute the effects of short-ranged impurities is a self-consistent Born approximation, $\Sigma=\gamma\, g^R_{0,0}(\omega)$, where $g^R_{0,0}(\omega)$ is the local Green function, see \cite{SM} for details.
Self-consistency is required to treat the regime where the Fermi energy is located inside the ultraflat bands. Importantly, within this method, one can take into account the formation of (disorder-broadened) LLs. The conductivity is calculated from the Kubo formula, without taking vertex corrections into account, see discussion below. We typically consider 200 bands (i.e., the lowest 100 LLs), allowing us to treat magnetic fields  $B\gtrsim 0.1\,$T. The effective disorder strength $\gamma \approx 0.06\, \text{nm}^3 (\text{eV})^2$ has been chosen to roughly reproduce the experimentally observed resistivities.

The effect of large-scale charge inhomogeneities is phenomenologically taken into account by averaging over the electronic density using a Gaussian  $p(n)\sim e^{-\frac{1}{2} (n-\bar n)^2/\sigma_n^2}$,   $\bar n=1.1 \times 10^{16} \text{ cm}^{-3}$, $\sigma_n \approx 0.3 \times 10^{16} \text{ cm}^{-3}$. The resistivies are highly anisotropic, $\rho_{bb} \gg \rho_{aa}, \rho_{cc}$. This affects the averaging in an inhomogeneous system: 
transport in the $b$ direction (in the $a$ direction) is effectively governed by a network of resistors connected in parallel (in series), respectively. Thus, $\rho_{bb} \approx (\int \sigma_{bb}(n) p(n) dn)^{-1}$ while
$\rho_{aa} \approx \int \rho_{aa}(n) p(n) dn$. The averaging over $n$ is needed
to explain the absence of clearly visible quantum oscillations [inset of Fig.~\ref{fig:Theory2}(d)] in the experiment.

In Fig.~\ref{fig:Theory2}(c) and \ref{fig:Theory2}(d), we show the resulting resistivities for $\boldsymbol {I} \parallel \boldsymbol {B}$ both in the $a$ and $b$ directions. While within our simplified model a full quantitative match cannot be achieved (see discussion below and in the supplementary material \cite{SM}), it captures the pronounced dip in the resistivity for $B\approx 2\,$T and $B\approx 0.2\,$T for currents in the $a$ and $b$ directions, respectively [see Fig. \ref{fig:exp2}(b) and \ref{fig:exp3}(a), respectively, for experimental data].

We can directly identify the origin of the dip. Upon increasing $B$, LLs form and the chemical potential $\mu$ becomes smaller and smaller. The dip occurs always in the regime, when the chemical potential is in the gray-shaded region shown in Fig.~\ref{fig:theory1}(b), where (i) only one LL contributes, and (ii) the band is strongly dispersive and not flat.

As we have shown in Sec. \ref{theo1}, the physics in this regime is quantitatively described by a chiral anomaly. 
%Originally introduced in high-energy physics, the chiral anomaly can be used to describe the negative magnetoresistance in Weyl systems. In its simplest version, a Weyl semimetal contains two Weyl points of opposite chirality. Denoting the density of electrons close to the Weyl-points by $\rho^{\pm}$, respectively, one obtains
%\begin{align}
%    \partial_t \rho^+-\partial_t \rho^- = \frac{e^3}{2 \pi^2 \hbar^2} \vec E \cdot \vec B -%\frac{\rho^+-\rho^-}{\tau}\label{anomalyMain}
%\end{align}
%where the $\vec E\cdot \vec B$ term with its universal prefactor describes the anomaly: parallel $\vec E$ and $\vec B$ fields effectively pump charges from one Weyl point to the next. The last term describes phenomenologically the effect of scattering by impurities. For large magnetic field the current in the Weyl system is given by $v_F (\rho^+-\rho^-)$. 
ZrTe$_5$ does {\em not} host any Weyl points but instead a single torus-shaped Fermi surface at zero magnetic field. Thus, Eq.~\eqref{main:anomaly} cannot be applied at small $B$. However, as we discussed in Sec. \ref{theo1}, the situation changes in the quantum regime,  where in the shaded region of Fig.~\ref{fig:theory1}(b) one can identify $\rho^\pm$ with the density of electrons moving with velocity $\pm v_F$. Here, the distance of the Fermi points in momentum space is set by the total width of the torus in direction parallel to the $B$ field.
Another feature of the chiral anomaly in Weyl semimetals is that the velocity $v_F$ is set by the dispersion of the Weyl node and thus approximately $B$-independent. Remarkably, this applies also to the nodal-line semimetal in the chiral-anomaly regime, see gray-shaded area in Fig.~\ref{fig:theory1}(b) and Fig.~\ref{fig:Theory2}.

%\YA{The reason why the chiral anomaly shows up as a dip, rather than an continuously negative magnetoresistance, lies in the magnetic-field dependence of $v_F$: Firstly, increasing $B$ brings $\mu$ down, which makes $v_F$ smaller as one can infer in Fig. 3(a); secondly, Eq.~\eqref{eq:gammaTot} says that smaller $v_F$ gives larger scattering rate, eventually increasing the resistivity.

 When the magnetic field is increased even further, the chemical potential enters the flat-band regime in Fig.~\ref{fig:theory1}(b). The scattering rate due to disorder, Fig.~\ref{fig:Theory2}(a), becomes
 much larger as it scales with $1/v_F$, see Eq.~\eqref{eq:gammaTot}.
 Thus, it gets larger than the bandwidth, violating Eq.~\eqref{eq:validity}, and a description of the system in terms of a chiral anomaly Eq.~\eqref{main:anomaly} is not possible any more. In this regime
the resistivity rises substantially, which explains why the negative magnetoresistance appears as a dip in our experiments.
For even larger fields, the numerical data in Fig.~\ref{fig:Theory2}(d) shows a drop of $\rho_{bb}$ for $B\gtrsim 1$\,T, which is not seen experimentally. In this regime localization physics is expected to become relevant, which is not described by our theory based on the self-consistent Born approximation. In the supplementary material \cite{SM}, we give an extended discussion of  the validity of our theory and show, for example, that it underestimates the dip in the resistivity in the chiral anomaly regime.

\section{Unusual planar Hall effect}

%\AR{WE HAVE CHANGED ALL XZ TO AB TO MAKE IT CONSISTENT WITH THE OTHER THEORY PARTS}
Now we show that the observed unusual PHE for $B$ in the $ab$-plane can be understood as a result of the magnetic-field-induced resistivity anisotropy due to the quasi-1D transport channel.
%in a similar way as the conventional PHE, even though the $\theta$-dependence is unconventional. 
%Since the chiral anomaly is irrelevant for $B \parallel ab$ and the formation of the quasi-1D channels in the quantum limit is the only source of the magnetic-field-induced anisotropy for this configuration, our result strongly suggests that the negative LMR in ZrTe$_5$ observed for $I \parallel a$ is also of this origin.
The key to understanding the unconventional $\theta$-dependence of the PHE is the strong anisotropy in the Fermi velocity $v_{\rm F}$ in ZrTe$_5$ \cite{Tang2019, Jiang2020, Wang2022}. 
%Without considering additional symmetry-breaking terms \cite{Wang2022}, the energy dispersion of the 3D Dirac band is $E_{\pm} = \sqrt{\hbar^2 (v_{{\rm F}, x}^2 k_x^2 + v_{{\rm F}, y}^2 k_y^2 + v_{{\rm F}, z}^2 k_z^2) +m^2 }$. 
Due to the van-der-Waals stacking along the $b$-axis, the $v_{\rm F}$ anisotropy is particularly strong between $a$ and $b$ axes, $v_a/v_b\approx 15.9$ \cite{Wang2022}. For the theoretical treatment of such an anisotropic system, one can rescale the axes to project to an isotropic system. For the discussion of the PHE in the $ab$-plane, it is sufficient to consider the rescaling in the $ab$ %(= $xz) 
plane. 
We try to rescale the coordinates and momentum to make the rescaled Fermi surface isotropic in the $ab$ plane. This can be done with the rescaling of coordinates $r_a' = r_a$, $r_b' = \lambda r_b$ and $k_a' = k_a$, $k_b' = k_b / \lambda$ with the scaling factor $\lambda \equiv v_{a}/v_{b}$. 
%We define $v_0 = v_{{\rm F}, x}= v_{{\rm F}, z} \lambda$. 
By defining the rescaling matrix 
\begin{equation} 
\boldsymbol{\lambda} \equiv
\left[\begin{array}{cc}
1 & 0 \\
0 & \lambda
\end{array}\right]\, , \label{eq2}
\end{equation}
we obtain the rescaling of the current density
$\boldsymbol{j}$ into $\boldsymbol{j'} = \boldsymbol{\lambda} \boldsymbol{j}$, the electric field $\boldsymbol{E}$ into $\boldsymbol{E'} = \boldsymbol{\lambda}^{-1} \boldsymbol{E}$, and, importantly, the 
magnetic field $\boldsymbol{B}$ into $\boldsymbol{B}'=B[\cos (\theta) / \lambda, \sin (\theta)]$ (see \cite{SM} for details). The effective magnetic-field angle $\theta'$ in the projected isotropic space is calculated with $\tan (\theta')=\lambda \cdot \tan (\theta)$. In general, the rescaling will also affect impurity potentials. If, however, the wavelength of electrons is larger than the width of impurity potentials, one may assume $\delta-$correlated disorder and the scattering is approximately isotropic in rescaled coordinates.

In the projected isotropic system, the effective magnetic field $\boldsymbol {B'}$ causes the formation of quasi-1D channels in the quantum limit (i.e. when all the electrons are condensed into the LLL). Since these quasi-1D channels are always along the direction of $\boldsymbol {B'}$, the transport is easier along $\boldsymbol {B'}$ and more difficult perpendicular to $\boldsymbol {B'}$ --- this is the origin of the magnetic-field-induced resistivity anisotropy. Indeed, the transverse magnetoresistance $\rho_{aa}(B)$ for $\boldsymbol {B} \parallel b$ shown in Fig.~\ref{fig:exp2}(b) presents a drastic increases from 4 m$\Omega$cm to 57 m$\Omega$cm as it reaches the quantum limit. Since the quasi-1D channels are always parallel to the magnetic field, when the direction of $\boldsymbol {B'}$ rotates in the $ab$ plane, the anisotropy axes of the resistivity tensor rotates together. (Note that we consider the case of sufficiently strong magnetic fields here; when the magnetic field is very weak, naturally the resistivity-anisotropy axes will not move.)

To simulate the $\theta$-dependence of the resistivity tensor, we make a crude assumption that in the projected isotropic system, the way the magnetic-field-induced resistivity anisotropy develops is independent of the orientation of $\boldsymbol {B'}$. This means that  the $B'$-dependencies of the resistivities along $\boldsymbol {B'}$ and perpendicular to $\boldsymbol {B'}$ do not change upon rotating $\boldsymbol {B'}$. With this assumption, we can use the experimentally-obtained $\rho_{aa}(B)$ and $\rho_{bb}(B)$ for $\boldsymbol {B} \parallel b$-axis ($\theta' = \theta$ = 90$^\circ$) as the input of the simulation, and calculate the rotated resistivity tensor (see \cite{SM} for details).
We chose $\boldsymbol {B} \parallel b$-axis for the reference measurement of $\rho_{aa}(B)$ and $\rho_{bb}(B)$, because the resistivity is stable with a slight tilt of $\boldsymbol {B}$ from $b$ [see Fig.~\ref{fig:exp2}(a)]; on the other hand, the resistivity changes a lot with a slight tilt of $\boldsymbol {B}$ from $a$, so $\boldsymbol {B} \parallel a$-axis is less suitable as a reference.

\begin{figure}[t]
	\centering
	\includegraphics[width=8.5cm]{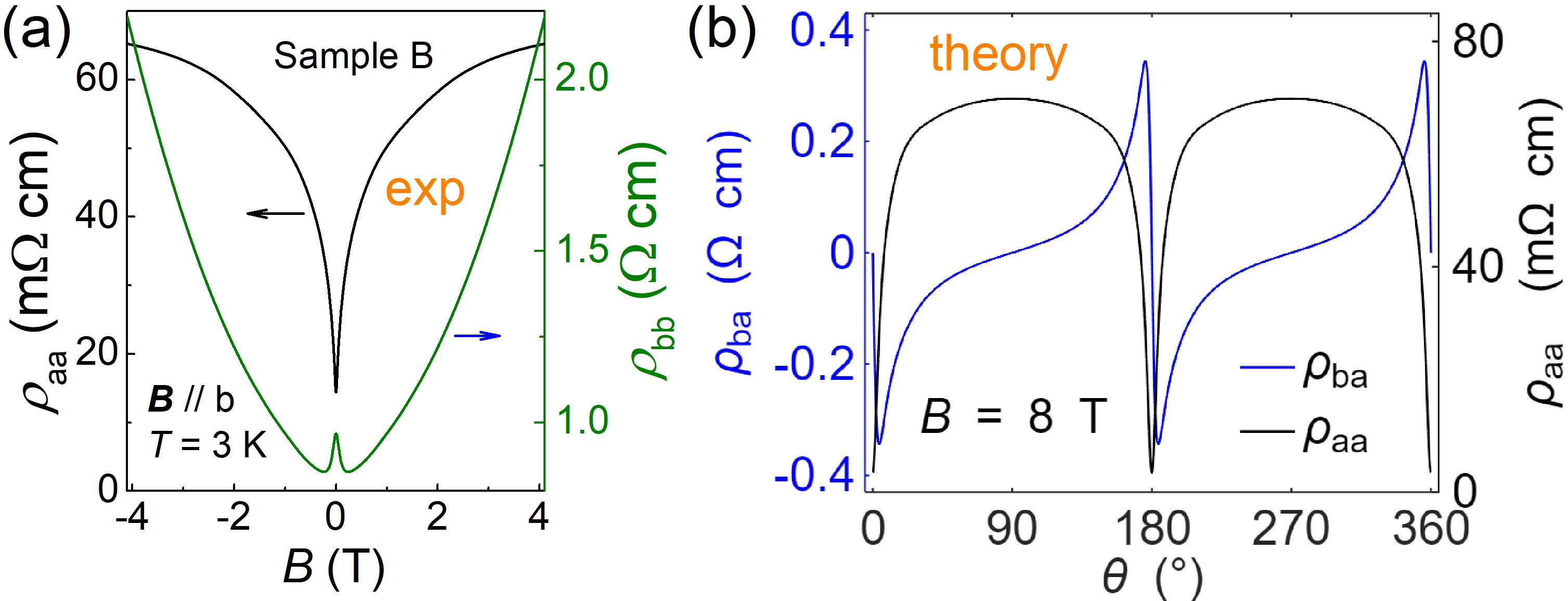}
	\caption{(a) $\rho_{aa}(B)$ and $\rho_{bb}(B)$ for $\boldsymbol {B} \parallel b$-axis measured in Sample B with the flux-transformer method \cite{SM}. (b) Angular dependencies of calculated $\rho_{ba}$ and $\rho_{aa}$ in 8 T; the calculations, based on Eq. \eqref{eq5}, used the data in (a) as the input.} \label{fig:exp3}
\end{figure}

To obtain the experimental input for the simulation, we measured $\rho_{aa}(B)$ and $\rho_{bb}(B)$ for $\theta = 90^{\circ}$ (i.e. $\boldsymbol {B} \parallel b$-axis) in sample B, which was prepared for the flux-transformer-type measurements of anisotropic resistivities \cite{Levin1997, Busch1992} (see Ref. \cite{SM} for details), and the data are shown in Fig.~\ref{fig:exp3}(a). One can see that $\rho_{bb}(B)$ presents a weak negative LMR at low field and $\rho_{aa}(B)$ presents a large singular TMR, both of which are essentially consistent with the behavior seen in Fig.~\ref{fig:exp2}(b) (note that the behavior of LMR is different between $\rho_{aa}$ and $\rho_{bb}$ near 0 T, but this is the field range outside of our consideration).  

After calculating the rotation of the rescaled frame, the resistivity tensor $\boldsymbol{\rho}(\theta, B)$ in the original unscaled system can be obtained as \cite{SM}
\small
\begin{equation}
\boldsymbol{\rho}(\theta, B)
= \left[\begin{array}{ll}
\Delta \rho \cdot \cos ^{2} \theta'+\rho_{a a}(B') & \lambda \Delta \rho \cdot \sin \theta' \cos \theta' \\
\lambda \Delta \rho \cdot \sin \theta' \cos \theta' & \lambda^2 \Delta \rho \cdot \sin ^{2} \theta' + \lambda^2 \rho_{a a}(B')
\end{array}\right], \label{eq5}
\end{equation}
\normalsize
where $\theta'= \tan^{-1} \left( \lambda \cdot \tan (\theta) \right)$, $\Delta \rho = \rho_{bb}(B') / \lambda^2 - \rho_{aa}(B')$ and $B'=B \sqrt{\cos ^{2} \theta / \lambda^{2}+\sin ^{2} \theta}$. 
We have calculated the theoretical $\theta$-dependent resistivity tensor obtained from Eq.~\eqref{eq5} using the data in Fig.~\ref{fig:exp3}(a), and the resulting $\rho_{aa}(\theta)$ and $\rho_{ba}(\theta)$ for $B$ = 8 T are shown in Fig.~\ref{fig:exp3}(b). 
It is remarkable that the peculiar $\theta$-dependencies of $\rho_{aa}$ and $\rho_{ba}$ in high magnetic fields, that are exemplified in Fig.~\ref{fig:exp2}(a), are essentially reproduced in this calculation. Thus, the unconventional $\theta$-dependence of the PHE is understood as a consequence of the large $v_{\rm F}$ anisotropy and the formation of quasi-1D channels that causes the magnetic-field-induced resistivity anisotropy.
Note that this calculation is based on the data from sample B, which has a larger $\rho_{aa}$ ($\approx$ 14.4 m$\Omega$cm) at 3 K in 0 T than that of sample A ($\approx$ 4.2 m$\Omega$cm). Thus, the calculated $\rho_{aa}$ and $\rho_{ba}$ shown in Fig.~\ref{fig:exp3}(b) are larger than those in Fig.~\ref{fig:exp2}(a).

\section{Summary and Conclusion}

Our analysis shows that the condensed-matter version of the chiral anomaly can generically take place in non-Weyl systems when the transport is dominated by the LLL yielding quasi-1D transport channels, as marked by the gray band in Fig.~\ref{fig:theory1}(b) and Fig.~\ref{fig:Theory2}. In this regime, the LLL is strongly dispersing with opposite sign of $v_F$ for $+k_F$ and $-k_F$; since the sign of $v_F$ defines the chirality in 1D transport, it is natural to view the imbalance between the $+k_F$ and $-k_F$ states of the LLL caused by the parallel electric field to be a chiral anomaly. We should note, however, that in the case of ZrTe$_5$, in a very high magnetic field where the chemical potential is brought down into the flat-band regime, scattering between electrons moving parallel and antiparallel to $\vec B$ becomes so strong that the chiral-anomaly physics gives way to localization physics. Our experimental data of the LMR in $\rho_{aa}$ and $\rho_{bb}$ in ZrTe$_5$ can be understood in this picture.

%The actual $B$-dependence of the LMR in ZrTe$_5$ is complicated due to the existence of charge puddles and the entrance of the flat-band physics at very high magnetic fields. Nevertheless, our theoretical calculations based on Born approximation can reproduce the key signature of the chiral anomaly in both $\rho_{aa}(B)$ and $\rho_{bb}(B)$. 

The quasi-1D transport channel in the LLL not only causes an enhanced conductance along $B$ (which is detected as the negative LMR), but also causes a reduced conductance for the transverse direction; namely, the formation of quasi one-dimensional channels leads to a strong magnetic-field-induced resistivity anisotropy, which is detected as the PHE. In the case of ZrTe$_5$, the strong anisotropy in $v_F$ gives rise to a complicated dependence of the resistivity anisotropy on the magnetic-field angle $\theta$ as described in Eq.~\eqref{eq5}. 
We showed that the unusual $\theta$-dependence of PHE in ZrTe$_5$ can actually be reconstructed from the experimentally-observed $\rho_{aa}(B)$ and $\rho_{bb}(B)$ with a simple rescaling argument.
This gives convincing evidence that the physics of the LLL that dictates the LMR in $\rho_{aa}$ and $\rho_{bb}$ lies behind the unusual PHE. 
In passing, we note that deviations from the standard $\sin(2\theta)$ behavior of PHE have been reported for (Ga,Mn)As \cite{Tang2003} and TaP \cite{Yang2019}, where it was caused by peculiar magnetic anisotropy \cite{Tang2003} or current jetting \cite{Yang2019}. The significant deviation from the standard behavior due to Fermi-velocity anisotropy is new and it demonstrates that PHE can be a powerful probe of the transport mechanism.

What are the conditions that the physics of the chiral anomaly results in pronounced transport signatures? One can argue \cite{Kikugawa2016} that  any sufficiently clean 3D system deep in the quantum limit exhibits a chiral anomaly. As in this regime the conductivity is proportional to $v_F^2$, Eq.~\eqref{eq:condVertex}, transport signatures depend, however, strongly on the value and the  $B$-dependence of $v_F$. Furthermore,
whether the system remains sufficiently weakly disordered for large $B$ also strongly depends on $v_F$, as the factor $\Gamma_{k_F}/v_F$ in Eq.~\eqref{eq:validity} is proportional to $1/v_F^2$.
Thus, Weyl systems where $v_F$ is constant and large are ideal systems to investigate transport signatures of the chiral anomaly. In contrast, in weakly doped insulators or semiconductors the value of $v_F$ is very small and drops with increasing $B$ due to the degeneracies of the Landau levels. This is, especially, an issue as the quantum limit can only be reached for large fields and small electron densities.
In the nodal line semimetal studied here, however,
the quantum limit is reached for small magnetic fields and we obtain clear transport signatures of the chiral anomaly without strong restrictions on sample quality. Furthermore, in the chiral anomaly regime, $v_F$ takes an approximately constant value, see Fig.~\ref{fig:Theory2}(b), which makes is  similar to the Weyl system in an intermediate field range.

%Besides, to the best of our knowledge this is the first time that a significant deviation from the standard $\sin(2\theta)$ behavior is observed for a PHE. It is interesting that the chiral anomaly can give rise not only to conventional PHE behavior \cite{Burkov2017, Nandy2017}, but also to such an unconventional PHE behavior. 

%Our theoretical model to explain the unusual $\theta$-dependence is a useful extension of the phenomenology of the PHE.
%By taking these LMR behavior along $a$- and $b$-axes, our simulation of the PHE can reproduce the unusual $\theta$-dependence. 
All in all, the present work establishes both theoretically and experimentally that the emergence of quasi-1D tranport channels in the LLL regime leads to a generic chiral anomaly in a non-Weyl system. This chiral anomaly primarily causes magnetic-field-induced resistivity anisotropy which results in negative LMR and a PHE. The negative LMR reported in the past for PdCoO$_2$ \cite{Kikugawa2016}, SrAs$_3$ \cite{Li2018b}, and ZrTe$_5$ \cite{Li2016} appears to fall into this category. It would be interesting to revisit unusual magnetotransport behavior in low-carrier-density systems
%\YA{having a nearly $B$-independent $v_F$} 
in the light of this generic chiral anomaly.

%Understood:\AR{We modified the discussion of the constant Fermi velocity as the previous version was not correct. While it is true that in our chiral anomaly regime we always the same value of the Fermi velocity, one cannot say that $v_F$ is constant as a function of $B$-field: we quickly leave the anomaly regime again.}

\vspace{3mm}
\acknowledgements{This work has received funding from the Deutsche Forschungsgemeinschaft (DFG, German Research Foundation) under CRC 1238-277146847 (subprojects A04, B01, and C02) and also from the DFG under Germany’s Excellence Strategy -- Cluster of Excellence Matter and Light for Quantum Computing (ML4Q) EXC 2004/1-390534769. We also acknowledge the support of the DFG Major instrumentation program under project No. 544410649.}
The raw data used in the generation of figures are available at the online repository Zenodo \cite{Zenodo}.

\bibliography{ZrTe5_bibliography}

\begin{thebibliography}{44}%
\makeatletter
\providecommand \@ifxundefined [1]{%
 \@ifx{#1\undefined}
}%
\providecommand \@ifnum [1]{%
 \ifnum #1\expandafter \@firstoftwo
 \else \expandafter \@secondoftwo
 \fi
}%
\providecommand \@ifx [1]{%
 \ifx #1\expandafter \@firstoftwo
 \else \expandafter \@secondoftwo
 \fi
}%
\providecommand \natexlab [1]{#1}%
\providecommand \enquote  [1]{``#1''}%
\providecommand \bibnamefont  [1]{#1}%
\providecommand \bibfnamefont [1]{#1}%
\providecommand \citenamefont [1]{#1}%
\providecommand \href@noop [0]{\@secondoftwo}%
\providecommand \href [0]{\begingroup \@sanitize@url \@href}%
\providecommand \@href[1]{\@@startlink{#1}\@@href}%
\providecommand \@@href[1]{\endgroup#1\@@endlink}%
\providecommand \@sanitize@url [0]{\catcode `\\12\catcode `\$12\catcode
  `\&12\catcode `\#12\catcode `\^12\catcode `\_12\catcode `\%12\relax}%
\providecommand \@@startlink[1]{}%
\providecommand \@@endlink[0]{}%
\providecommand \url  [0]{\begingroup\@sanitize@url \@url }%
\providecommand \@url [1]{\endgroup\@href {#1}{\urlprefix }}%
\providecommand \urlprefix  [0]{URL }%
\providecommand \Eprint [0]{\href }%
\providecommand \doibase [0]{https://doi.org/}%
\providecommand \selectlanguage [0]{\@gobble}%
\providecommand \bibinfo  [0]{\@secondoftwo}%
\providecommand \bibfield  [0]{\@secondoftwo}%
\providecommand \translation [1]{[#1]}%
\providecommand \BibitemOpen [0]{}%
\providecommand \bibitemStop [0]{}%
\providecommand \bibitemNoStop [0]{.\EOS\space}%
\providecommand \EOS [0]{\spacefactor3000\relax}%
\providecommand \BibitemShut  [1]{\csname bibitem#1\endcsname}%
\let\auto@bib@innerbib\@empty
%</preamble>
\bibitem [{\citenamefont {Armitage}\ \emph {et~al.}(2018)\citenamefont
  {Armitage}, \citenamefont {Mele},\ and\ \citenamefont
  {Vishwanath}}]{Armitage2018}%
  \BibitemOpen
  \bibfield  {author} {\bibinfo {author} {\bibfnamefont {N.}~\bibnamefont
  {Armitage}}, \bibinfo {author} {\bibfnamefont {E.~J.}\ \bibnamefont {Mele}},\
  and\ \bibinfo {author} {\bibfnamefont {A.}~\bibnamefont {Vishwanath}},\
  }\bibfield  {title} {\bibinfo {title} {{W}eyl and {D}irac semimetals in
  three-dimensional solids},\ }\href@noop {} {\bibfield  {journal} {\bibinfo
  {journal} {Rev. Mod. Phys.}\ }\textbf {\bibinfo {volume} {90}},\ \bibinfo
  {pages} {015001} (\bibinfo {year} {2018})}\BibitemShut {NoStop}%
\bibitem [{\citenamefont {Nielsen}\ and\ \citenamefont
  {Ninomiya}(1983)}]{Nielsen1983}%
  \BibitemOpen
  \bibfield  {author} {\bibinfo {author} {\bibfnamefont {H.~B.}\ \bibnamefont
  {Nielsen}}\ and\ \bibinfo {author} {\bibfnamefont {M.}~\bibnamefont
  {Ninomiya}},\ }\bibfield  {title} {\bibinfo {title} {The {Adler-Bell-Jackiw
  anomaly and Weyl} fermions in a crystal},\ }\href
  {https://doi.org/http://dx.doi.org/10.1016/0370-2693(83)91529-0} {\bibfield
  {journal} {\bibinfo  {journal} {Phys. Lett. B}\ }\textbf {\bibinfo {volume}
  {130}},\ \bibinfo {pages} {389} (\bibinfo {year} {1983})}\BibitemShut
  {NoStop}%
\bibitem [{\citenamefont {Son}\ and\ \citenamefont {Spivak}(2013)}]{Son2013}%
  \BibitemOpen
  \bibfield  {author} {\bibinfo {author} {\bibfnamefont {D.}~\bibnamefont
  {Son}}\ and\ \bibinfo {author} {\bibfnamefont {B.}~\bibnamefont {Spivak}},\
  }\bibfield  {title} {\bibinfo {title} {Chiral anomaly and classical negative
  magnetoresistance of weyl metals},\ }\href@noop {} {\bibfield  {journal}
  {\bibinfo  {journal} {Phys. Rev. B}\ }\textbf {\bibinfo {volume} {88}},\
  \bibinfo {pages} {104412} (\bibinfo {year} {2013})}\BibitemShut {NoStop}%
\bibitem [{\citenamefont {Burkov}(2017)}]{Burkov2017}%
  \BibitemOpen
  \bibfield  {author} {\bibinfo {author} {\bibfnamefont {A.}~\bibnamefont
  {Burkov}},\ }\bibfield  {title} {\bibinfo {title} {Giant planar {H}all effect
  in topological metals},\ }\href@noop {} {\bibfield  {journal} {\bibinfo
  {journal} {Phys. Rev. B}\ }\textbf {\bibinfo {volume} {96}},\ \bibinfo
  {pages} {041110} (\bibinfo {year} {2017})}\BibitemShut {NoStop}%
\bibitem [{\citenamefont {Nandy}\ \emph {et~al.}(2017)\citenamefont {Nandy},
  \citenamefont {Sharma}, \citenamefont {Taraphder},\ and\ \citenamefont
  {Tewari}}]{Nandy2017}%
  \BibitemOpen
  \bibfield  {author} {\bibinfo {author} {\bibfnamefont {S.}~\bibnamefont
  {Nandy}}, \bibinfo {author} {\bibfnamefont {G.}~\bibnamefont {Sharma}},
  \bibinfo {author} {\bibfnamefont {A.}~\bibnamefont {Taraphder}},\ and\
  \bibinfo {author} {\bibfnamefont {S.}~\bibnamefont {Tewari}},\ }\bibfield
  {title} {\bibinfo {title} {Chiral {A}nomaly as the {O}rigin of the {P}lanar
  {H}all {E}ffect in {W}eyl {S}emimetals},\ }\href@noop {} {\bibfield
  {journal} {\bibinfo  {journal} {Phys. Rev. Lett.}\ }\textbf {\bibinfo
  {volume} {119}},\ \bibinfo {pages} {176804} (\bibinfo {year}
  {2017})}\BibitemShut {NoStop}%
\bibitem [{\citenamefont {Huang}\ \emph {et~al.}(2015)\citenamefont {Huang},
  \citenamefont {Zhao}, \citenamefont {Long}, \citenamefont {Wang},
  \citenamefont {Chen}, \citenamefont {Yang}, \citenamefont {Liang},
  \citenamefont {Xue}, \citenamefont {Weng}, \citenamefont {Fang} \emph
  {et~al.}}]{Huang2015}%
  \BibitemOpen
  \bibfield  {author} {\bibinfo {author} {\bibfnamefont {X.}~\bibnamefont
  {Huang}}, \bibinfo {author} {\bibfnamefont {L.}~\bibnamefont {Zhao}},
  \bibinfo {author} {\bibfnamefont {Y.}~\bibnamefont {Long}}, \bibinfo {author}
  {\bibfnamefont {P.}~\bibnamefont {Wang}}, \bibinfo {author} {\bibfnamefont
  {D.}~\bibnamefont {Chen}}, \bibinfo {author} {\bibfnamefont {Z.}~\bibnamefont
  {Yang}}, \bibinfo {author} {\bibfnamefont {H.}~\bibnamefont {Liang}},
  \bibinfo {author} {\bibfnamefont {M.}~\bibnamefont {Xue}}, \bibinfo {author}
  {\bibfnamefont {H.}~\bibnamefont {Weng}}, \bibinfo {author} {\bibfnamefont
  {Z.}~\bibnamefont {Fang}}, \emph {et~al.},\ }\bibfield  {title} {\bibinfo
  {title} {Observation of the chiral-anomaly-induced negative magnetoresistance
  in 3{D} {W}eyl semimetal {TaAs}},\ }\href@noop {} {\bibfield  {journal}
  {\bibinfo  {journal} {Phys. Rev. X}\ }\textbf {\bibinfo {volume} {5}},\
  \bibinfo {pages} {031023} (\bibinfo {year} {2015})}\BibitemShut {NoStop}%
\bibitem [{\citenamefont {Zhang}\ \emph {et~al.}(2016)\citenamefont {Zhang},
  \citenamefont {Xu}, \citenamefont {Belopolski}, \citenamefont {Yuan},
  \citenamefont {Lin}, \citenamefont {Tong}, \citenamefont {Bian},
  \citenamefont {Alidoust}, \citenamefont {Lee}, \citenamefont {Huang} \emph
  {et~al.}}]{Zhang2016}%
  \BibitemOpen
  \bibfield  {author} {\bibinfo {author} {\bibfnamefont {C.-L.}\ \bibnamefont
  {Zhang}}, \bibinfo {author} {\bibfnamefont {S.-Y.}\ \bibnamefont {Xu}},
  \bibinfo {author} {\bibfnamefont {I.}~\bibnamefont {Belopolski}}, \bibinfo
  {author} {\bibfnamefont {Z.}~\bibnamefont {Yuan}}, \bibinfo {author}
  {\bibfnamefont {Z.}~\bibnamefont {Lin}}, \bibinfo {author} {\bibfnamefont
  {B.}~\bibnamefont {Tong}}, \bibinfo {author} {\bibfnamefont {G.}~\bibnamefont
  {Bian}}, \bibinfo {author} {\bibfnamefont {N.}~\bibnamefont {Alidoust}},
  \bibinfo {author} {\bibfnamefont {C.-C.}\ \bibnamefont {Lee}}, \bibinfo
  {author} {\bibfnamefont {S.-M.}\ \bibnamefont {Huang}}, \emph {et~al.},\
  }\bibfield  {title} {\bibinfo {title} {Signatures of the
  {A}dler-{B}ell-{J}ackiw chiral anomaly in a {W}eyl fermion semimetal},\
  }\href@noop {} {\bibfield  {journal} {\bibinfo  {journal} {Nat. Commun.}\
  }\textbf {\bibinfo {volume} {7}},\ \bibinfo {pages} {1} (\bibinfo {year}
  {2016})}\BibitemShut {NoStop}%
\bibitem [{\citenamefont {Wang}\ \emph
  {et~al.}(2016{\natexlab{a}})\citenamefont {Wang}, \citenamefont {Zheng},
  \citenamefont {Shen}, \citenamefont {Lu}, \citenamefont {Fang}, \citenamefont
  {Sheng}, \citenamefont {Zhou}, \citenamefont {Yang}, \citenamefont {Li},
  \citenamefont {Feng} \emph {et~al.}}]{Wang2016}%
  \BibitemOpen
  \bibfield  {author} {\bibinfo {author} {\bibfnamefont {Z.}~\bibnamefont
  {Wang}}, \bibinfo {author} {\bibfnamefont {Y.}~\bibnamefont {Zheng}},
  \bibinfo {author} {\bibfnamefont {Z.}~\bibnamefont {Shen}}, \bibinfo {author}
  {\bibfnamefont {Y.}~\bibnamefont {Lu}}, \bibinfo {author} {\bibfnamefont
  {H.}~\bibnamefont {Fang}}, \bibinfo {author} {\bibfnamefont {F.}~\bibnamefont
  {Sheng}}, \bibinfo {author} {\bibfnamefont {Y.}~\bibnamefont {Zhou}},
  \bibinfo {author} {\bibfnamefont {X.}~\bibnamefont {Yang}}, \bibinfo {author}
  {\bibfnamefont {Y.}~\bibnamefont {Li}}, \bibinfo {author} {\bibfnamefont
  {C.}~\bibnamefont {Feng}}, \emph {et~al.},\ }\bibfield  {title} {\bibinfo
  {title} {Helicity-protected ultrahigh mobility {W}eyl fermions in {NbP}},\
  }\href@noop {} {\bibfield  {journal} {\bibinfo  {journal} {Phys. Rev. B}\
  }\textbf {\bibinfo {volume} {93}},\ \bibinfo {pages} {121112} (\bibinfo
  {year} {2016}{\natexlab{a}})}\BibitemShut {NoStop}%
\bibitem [{\citenamefont {Arnold}\ \emph {et~al.}(2016)\citenamefont {Arnold},
  \citenamefont {Shekhar}, \citenamefont {Wu}, \citenamefont {Sun},
  \citenamefont {dos Reis}, \citenamefont {Kumar}, \citenamefont {Naumann},
  \citenamefont {Ajeesh}, \citenamefont {Schmidt}, \citenamefont {Grushin},
  \citenamefont {Bardarson}, \citenamefont {Baenitz}, \citenamefont {Sokolov},
  \citenamefont {Borrmann}, \citenamefont {Nicklas}, \citenamefont {Felser},
  \citenamefont {Hassinger},\ and\ \citenamefont {Yan}}]{Arnold2016}%
  \BibitemOpen
  \bibfield  {author} {\bibinfo {author} {\bibfnamefont {F.}~\bibnamefont
  {Arnold}}, \bibinfo {author} {\bibfnamefont {C.}~\bibnamefont {Shekhar}},
  \bibinfo {author} {\bibfnamefont {S.-C.}\ \bibnamefont {Wu}}, \bibinfo
  {author} {\bibfnamefont {Y.}~\bibnamefont {Sun}}, \bibinfo {author}
  {\bibfnamefont {R.~D.}\ \bibnamefont {dos Reis}}, \bibinfo {author}
  {\bibfnamefont {N.}~\bibnamefont {Kumar}}, \bibinfo {author} {\bibfnamefont
  {M.}~\bibnamefont {Naumann}}, \bibinfo {author} {\bibfnamefont {M.~O.}\
  \bibnamefont {Ajeesh}}, \bibinfo {author} {\bibfnamefont {M.}~\bibnamefont
  {Schmidt}}, \bibinfo {author} {\bibfnamefont {A.~G.}\ \bibnamefont
  {Grushin}}, \bibinfo {author} {\bibfnamefont {J.~H.}\ \bibnamefont
  {Bardarson}}, \bibinfo {author} {\bibfnamefont {M.}~\bibnamefont {Baenitz}},
  \bibinfo {author} {\bibfnamefont {D.}~\bibnamefont {Sokolov}}, \bibinfo
  {author} {\bibfnamefont {H.}~\bibnamefont {Borrmann}}, \bibinfo {author}
  {\bibfnamefont {M.}~\bibnamefont {Nicklas}}, \bibinfo {author} {\bibfnamefont
  {C.}~\bibnamefont {Felser}}, \bibinfo {author} {\bibfnamefont
  {E.}~\bibnamefont {Hassinger}},\ and\ \bibinfo {author} {\bibfnamefont
  {B.}~\bibnamefont {Yan}},\ }\bibfield  {title} {\bibinfo {title} {Negative
  magnetoresistance without well-defined chirality in the {Weyl} semimetal
  {TaP}},\ }\href {https://doi.org/10.1038/ncomms11615} {\bibfield  {journal}
  {\bibinfo  {journal} {Nat. Commun.}\ }\textbf {\bibinfo {volume} {7}},\
  \bibinfo {pages} {11615} (\bibinfo {year} {2016})}\BibitemShut {NoStop}%
\bibitem [{\citenamefont {Wang}\ \emph
  {et~al.}(2016{\natexlab{b}})\citenamefont {Wang}, \citenamefont {Liu},
  \citenamefont {Liu}, \citenamefont {Pan}, \citenamefont {Zhang},
  \citenamefont {Zeng}, \citenamefont {Fu}, \citenamefont {Wang}, \citenamefont
  {Xu}, \citenamefont {Huang} \emph {et~al.}}]{Wang2016a}%
  \BibitemOpen
  \bibfield  {author} {\bibinfo {author} {\bibfnamefont {Y.}~\bibnamefont
  {Wang}}, \bibinfo {author} {\bibfnamefont {E.}~\bibnamefont {Liu}}, \bibinfo
  {author} {\bibfnamefont {H.}~\bibnamefont {Liu}}, \bibinfo {author}
  {\bibfnamefont {Y.}~\bibnamefont {Pan}}, \bibinfo {author} {\bibfnamefont
  {L.}~\bibnamefont {Zhang}}, \bibinfo {author} {\bibfnamefont
  {J.}~\bibnamefont {Zeng}}, \bibinfo {author} {\bibfnamefont {Y.}~\bibnamefont
  {Fu}}, \bibinfo {author} {\bibfnamefont {M.}~\bibnamefont {Wang}}, \bibinfo
  {author} {\bibfnamefont {K.}~\bibnamefont {Xu}}, \bibinfo {author}
  {\bibfnamefont {Z.}~\bibnamefont {Huang}}, \emph {et~al.},\ }\bibfield
  {title} {\bibinfo {title} {Gate-tunable negative longitudinal
  magnetoresistance in the predicted type-{II} {W}eyl semimetal {WTe$_2$}},\
  }\href@noop {} {\bibfield  {journal} {\bibinfo  {journal} {Nat. Commun.}\
  }\textbf {\bibinfo {volume} {7}},\ \bibinfo {pages} {13142} (\bibinfo {year}
  {2016}{\natexlab{b}})}\BibitemShut {NoStop}%
\bibitem [{\citenamefont {Liu}\ \emph {et~al.}(2018)\citenamefont {Liu},
  \citenamefont {Sun}, \citenamefont {Kumar}, \citenamefont {Muechler},
  \citenamefont {Sun}, \citenamefont {Jiao}, \citenamefont {Yang},
  \citenamefont {Liu}, \citenamefont {Liang}, \citenamefont {Xu} \emph
  {et~al.}}]{Liu2018}%
  \BibitemOpen
  \bibfield  {author} {\bibinfo {author} {\bibfnamefont {E.}~\bibnamefont
  {Liu}}, \bibinfo {author} {\bibfnamefont {Y.}~\bibnamefont {Sun}}, \bibinfo
  {author} {\bibfnamefont {N.}~\bibnamefont {Kumar}}, \bibinfo {author}
  {\bibfnamefont {L.}~\bibnamefont {Muechler}}, \bibinfo {author}
  {\bibfnamefont {A.}~\bibnamefont {Sun}}, \bibinfo {author} {\bibfnamefont
  {L.}~\bibnamefont {Jiao}}, \bibinfo {author} {\bibfnamefont {S.-Y.}\
  \bibnamefont {Yang}}, \bibinfo {author} {\bibfnamefont {D.}~\bibnamefont
  {Liu}}, \bibinfo {author} {\bibfnamefont {A.}~\bibnamefont {Liang}}, \bibinfo
  {author} {\bibfnamefont {Q.}~\bibnamefont {Xu}}, \emph {et~al.},\ }\bibfield
  {title} {\bibinfo {title} {Giant anomalous {H}all effect in a ferromagnetic
  kagome-lattice semimetal},\ }\href@noop {} {\bibfield  {journal} {\bibinfo
  {journal} {Nat. Phys.}\ }\textbf {\bibinfo {volume} {14}},\ \bibinfo {pages}
  {1125} (\bibinfo {year} {2018})}\BibitemShut {NoStop}%
\bibitem [{\citenamefont {Xiong}\ \emph {et~al.}(2015)\citenamefont {Xiong},
  \citenamefont {Kushwaha}, \citenamefont {Liang}, \citenamefont {Krizan},
  \citenamefont {Hirschberger}, \citenamefont {Wang}, \citenamefont {Cava},\
  and\ \citenamefont {Ong}}]{Xiong2015}%
  \BibitemOpen
  \bibfield  {author} {\bibinfo {author} {\bibfnamefont {J.}~\bibnamefont
  {Xiong}}, \bibinfo {author} {\bibfnamefont {S.~K.}\ \bibnamefont {Kushwaha}},
  \bibinfo {author} {\bibfnamefont {T.}~\bibnamefont {Liang}}, \bibinfo
  {author} {\bibfnamefont {J.~W.}\ \bibnamefont {Krizan}}, \bibinfo {author}
  {\bibfnamefont {M.}~\bibnamefont {Hirschberger}}, \bibinfo {author}
  {\bibfnamefont {W.}~\bibnamefont {Wang}}, \bibinfo {author} {\bibfnamefont
  {R.~J.}\ \bibnamefont {Cava}},\ and\ \bibinfo {author} {\bibfnamefont
  {N.~P.}\ \bibnamefont {Ong}},\ }\bibfield  {title} {\bibinfo {title}
  {Evidence for the chiral anomaly in the {D}irac semimetal {Na$_3$Bi}},\
  }\href@noop {} {\bibfield  {journal} {\bibinfo  {journal} {Science}\ }\textbf
  {\bibinfo {volume} {350}},\ \bibinfo {pages} {413} (\bibinfo {year}
  {2015})}\BibitemShut {NoStop}%
\bibitem [{\citenamefont {Hirschberger}\ \emph {et~al.}(2016)\citenamefont
  {Hirschberger}, \citenamefont {Kushwaha}, \citenamefont {Wang}, \citenamefont
  {Gibson}, \citenamefont {Liang}, \citenamefont {Belvin}, \citenamefont
  {Bernevig}, \citenamefont {Cava},\ and\ \citenamefont
  {Ong}}]{Hirschberger2016}%
  \BibitemOpen
  \bibfield  {author} {\bibinfo {author} {\bibfnamefont {M.}~\bibnamefont
  {Hirschberger}}, \bibinfo {author} {\bibfnamefont {S.}~\bibnamefont
  {Kushwaha}}, \bibinfo {author} {\bibfnamefont {Z.}~\bibnamefont {Wang}},
  \bibinfo {author} {\bibfnamefont {Q.}~\bibnamefont {Gibson}}, \bibinfo
  {author} {\bibfnamefont {S.}~\bibnamefont {Liang}}, \bibinfo {author}
  {\bibfnamefont {C.~A.}\ \bibnamefont {Belvin}}, \bibinfo {author}
  {\bibfnamefont {B.~A.}\ \bibnamefont {Bernevig}}, \bibinfo {author}
  {\bibfnamefont {R.~J.}\ \bibnamefont {Cava}},\ and\ \bibinfo {author}
  {\bibfnamefont {N.~P.}\ \bibnamefont {Ong}},\ }\bibfield  {title} {\bibinfo
  {title} {The chiral anomaly and thermopower of {W}eyl fermions in the
  half-{H}eusler {GdPtBi}},\ }\href@noop {} {\bibfield  {journal} {\bibinfo
  {journal} {Nat. Mater.}\ }\textbf {\bibinfo {volume} {15}},\ \bibinfo {pages}
  {1161} (\bibinfo {year} {2016})}\BibitemShut {NoStop}%
\bibitem [{\citenamefont {Li}\ \emph {et~al.}(2015)\citenamefont {Li},
  \citenamefont {Wang}, \citenamefont {Liu}, \citenamefont {Wang},
  \citenamefont {Liao},\ and\ \citenamefont {Yu}}]{Li2015}%
  \BibitemOpen
  \bibfield  {author} {\bibinfo {author} {\bibfnamefont {C.-Z.}\ \bibnamefont
  {Li}}, \bibinfo {author} {\bibfnamefont {L.-X.}\ \bibnamefont {Wang}},
  \bibinfo {author} {\bibfnamefont {H.}~\bibnamefont {Liu}}, \bibinfo {author}
  {\bibfnamefont {J.}~\bibnamefont {Wang}}, \bibinfo {author} {\bibfnamefont
  {Z.-M.}\ \bibnamefont {Liao}},\ and\ \bibinfo {author} {\bibfnamefont
  {D.-P.}\ \bibnamefont {Yu}},\ }\bibfield  {title} {\bibinfo {title} {Giant
  negative magnetoresistance induced by the chiral anomaly in individual
  {Cd$_3$As$_2$} nanowires},\ }\href@noop {} {\bibfield  {journal} {\bibinfo
  {journal} {Nat. Commun.}\ }\textbf {\bibinfo {volume} {6}},\ \bibinfo {pages}
  {10137} (\bibinfo {year} {2015})}\BibitemShut {NoStop}%
\bibitem [{\citenamefont {Li}\ \emph {et~al.}(2016{\natexlab{a}})\citenamefont
  {Li}, \citenamefont {He}, \citenamefont {Lu}, \citenamefont {Zhang},
  \citenamefont {Liu}, \citenamefont {Ma}, \citenamefont {Fan}, \citenamefont
  {Shen},\ and\ \citenamefont {Wang}}]{Li2016a}%
  \BibitemOpen
  \bibfield  {author} {\bibinfo {author} {\bibfnamefont {H.}~\bibnamefont
  {Li}}, \bibinfo {author} {\bibfnamefont {H.}~\bibnamefont {He}}, \bibinfo
  {author} {\bibfnamefont {H.-Z.}\ \bibnamefont {Lu}}, \bibinfo {author}
  {\bibfnamefont {H.}~\bibnamefont {Zhang}}, \bibinfo {author} {\bibfnamefont
  {H.}~\bibnamefont {Liu}}, \bibinfo {author} {\bibfnamefont {R.}~\bibnamefont
  {Ma}}, \bibinfo {author} {\bibfnamefont {Z.}~\bibnamefont {Fan}}, \bibinfo
  {author} {\bibfnamefont {S.-Q.}\ \bibnamefont {Shen}},\ and\ \bibinfo
  {author} {\bibfnamefont {J.}~\bibnamefont {Wang}},\ }\bibfield  {title}
  {\bibinfo {title} {Negative magnetoresistance in {D}irac semimetal
  {Cd$_3$As$_2$}},\ }\href@noop {} {\bibfield  {journal} {\bibinfo  {journal}
  {Nat. Commun.}\ }\textbf {\bibinfo {volume} {7}},\ \bibinfo {pages} {10301}
  (\bibinfo {year} {2016}{\natexlab{a}})}\BibitemShut {NoStop}%
\bibitem [{\citenamefont {Ong}\ and\ \citenamefont {Liang}(2021)}]{Ong2021}%
  \BibitemOpen
  \bibfield  {author} {\bibinfo {author} {\bibfnamefont {N.~P.}\ \bibnamefont
  {Ong}}\ and\ \bibinfo {author} {\bibfnamefont {S.}~\bibnamefont {Liang}},\
  }\bibfield  {title} {\bibinfo {title} {Experimental signatures of the chiral
  anomaly in {D}irac–{W}eyl semimetals},\ }\href
  {https://doi.org/10.1038/s42254-021-00310-9} {\bibfield  {journal} {\bibinfo
  {journal} {Nat. Rev. Phys.}\ }\textbf {\bibinfo {volume} {3}},\ \bibinfo
  {pages} {394} (\bibinfo {year} {2021})}\BibitemShut {NoStop}%
\bibitem [{\citenamefont {Li}\ \emph {et~al.}(2016{\natexlab{b}})\citenamefont
  {Li}, \citenamefont {Kharzeev}, \citenamefont {Zhang}, \citenamefont {Huang},
  \citenamefont {Pletikosic}, \citenamefont {Fedorov}, \citenamefont {Zhong},
  \citenamefont {Schneeloch}, \citenamefont {Gu},\ and\ \citenamefont
  {Valla}}]{Li2016}%
  \BibitemOpen
  \bibfield  {author} {\bibinfo {author} {\bibfnamefont {Q.}~\bibnamefont
  {Li}}, \bibinfo {author} {\bibfnamefont {D.~E.}\ \bibnamefont {Kharzeev}},
  \bibinfo {author} {\bibfnamefont {C.}~\bibnamefont {Zhang}}, \bibinfo
  {author} {\bibfnamefont {Y.}~\bibnamefont {Huang}}, \bibinfo {author}
  {\bibfnamefont {I.}~\bibnamefont {Pletikosic}}, \bibinfo {author}
  {\bibfnamefont {A.~V.}\ \bibnamefont {Fedorov}}, \bibinfo {author}
  {\bibfnamefont {R.~D.}\ \bibnamefont {Zhong}}, \bibinfo {author}
  {\bibfnamefont {J.~A.}\ \bibnamefont {Schneeloch}}, \bibinfo {author}
  {\bibfnamefont {G.~D.}\ \bibnamefont {Gu}},\ and\ \bibinfo {author}
  {\bibfnamefont {T.}~\bibnamefont {Valla}},\ }\bibfield  {title} {\bibinfo
  {title} {Chiral magnetic effect in {Z}r{T}e$_5$},\ }\href
  {https://doi.org/10.1038/nphys3648
  http://www.nature.com/nphys/journal/v12/n6/abs/nphys3648.html#supplementary-information}
  {\bibfield  {journal} {\bibinfo  {journal} {Nat. Phys.}\ }\textbf {\bibinfo
  {volume} {12}},\ \bibinfo {pages} {550} (\bibinfo {year}
  {2016}{\natexlab{b}})}\BibitemShut {NoStop}%
\bibitem [{\citenamefont {Weng}\ \emph {et~al.}(2014)\citenamefont {Weng},
  \citenamefont {Dai},\ and\ \citenamefont {Fang}}]{Weng2014}%
  \BibitemOpen
  \bibfield  {author} {\bibinfo {author} {\bibfnamefont {H.}~\bibnamefont
  {Weng}}, \bibinfo {author} {\bibfnamefont {X.}~\bibnamefont {Dai}},\ and\
  \bibinfo {author} {\bibfnamefont {Z.}~\bibnamefont {Fang}},\ }\bibfield
  {title} {\bibinfo {title} {Transition-metal {P}entatelluride {Z}r{T}e$_5$ and
  {H}f{T}e$_5$: {A} {P}aradigm for {L}arge-{G}ap {Q}uantum {S}pin {H}all
  {I}nsulators},\ }\href@noop {} {\bibfield  {journal} {\bibinfo  {journal}
  {Phys. Rev. X}\ }\textbf {\bibinfo {volume} {4}},\ \bibinfo {pages} {011002}
  (\bibinfo {year} {2014})}\BibitemShut {NoStop}%
\bibitem [{\citenamefont {Chen}\ \emph {et~al.}(2015)\citenamefont {Chen},
  \citenamefont {Chen}, \citenamefont {Song}, \citenamefont {Schneeloch},
  \citenamefont {Gu}, \citenamefont {Wang},\ and\ \citenamefont
  {Wang}}]{RYChen2015}%
  \BibitemOpen
  \bibfield  {author} {\bibinfo {author} {\bibfnamefont {R.~Y.}\ \bibnamefont
  {Chen}}, \bibinfo {author} {\bibfnamefont {Z.~G.}\ \bibnamefont {Chen}},
  \bibinfo {author} {\bibfnamefont {X.~Y.}\ \bibnamefont {Song}}, \bibinfo
  {author} {\bibfnamefont {J.~A.}\ \bibnamefont {Schneeloch}}, \bibinfo
  {author} {\bibfnamefont {G.~D.}\ \bibnamefont {Gu}}, \bibinfo {author}
  {\bibfnamefont {F.}~\bibnamefont {Wang}},\ and\ \bibinfo {author}
  {\bibfnamefont {N.~L.}\ \bibnamefont {Wang}},\ }\bibfield  {title} {\bibinfo
  {title} {Magnetoinfrared {S}pectroscopy of {L}andau {L}evels and {Z}eeman
  {S}plitting of {T}hree-{D}imensional {M}assless {D}irac {F}ermions in
  {Z}r{T}e$_5$},\ }\href {https://doi.org/10.1103/PhysRevLett.115.176404}
  {\bibfield  {journal} {\bibinfo  {journal} {Phys. Rev. Lett.}\ }\textbf
  {\bibinfo {volume} {115}},\ \bibinfo {pages} {176404} (\bibinfo {year}
  {2015})}\BibitemShut {NoStop}%
\bibitem [{\citenamefont {Wang}\ \emph {et~al.}(2022)\citenamefont {Wang},
  \citenamefont {Legg}, \citenamefont {B{\"o}merich}, \citenamefont {Park},
  \citenamefont {Biesenkamp}, \citenamefont {Taskin}, \citenamefont {Braden},
  \citenamefont {Rosch},\ and\ \citenamefont {Ando}}]{Wang2022}%
  \BibitemOpen
  \bibfield  {author} {\bibinfo {author} {\bibfnamefont {Y.}~\bibnamefont
  {Wang}}, \bibinfo {author} {\bibfnamefont {H.~F.}\ \bibnamefont {Legg}},
  \bibinfo {author} {\bibfnamefont {T.}~\bibnamefont {B{\"o}merich}}, \bibinfo
  {author} {\bibfnamefont {J.}~\bibnamefont {Park}}, \bibinfo {author}
  {\bibfnamefont {S.}~\bibnamefont {Biesenkamp}}, \bibinfo {author}
  {\bibfnamefont {A.}~\bibnamefont {Taskin}}, \bibinfo {author} {\bibfnamefont
  {M.}~\bibnamefont {Braden}}, \bibinfo {author} {\bibfnamefont
  {A.}~\bibnamefont {Rosch}},\ and\ \bibinfo {author} {\bibfnamefont
  {Y.}~\bibnamefont {Ando}},\ }\bibfield  {title} {\bibinfo {title} {Gigantic
  {M}agnetochiral {A}nisotropy in the {T}opological {S}emimetal {ZrTe$_5$}},\
  }\href@noop {} {\bibfield  {journal} {\bibinfo  {journal} {Phys. Rev. Lett.}\
  }\textbf {\bibinfo {volume} {128}},\ \bibinfo {pages} {176602} (\bibinfo
  {year} {2022})}\BibitemShut {NoStop}%
\bibitem [{\citenamefont {Wang}\ \emph {et~al.}(2023)\citenamefont {Wang},
  \citenamefont {B{\"o}merich}, \citenamefont {Park}, \citenamefont {Legg},
  \citenamefont {Taskin}, \citenamefont {Rosch},\ and\ \citenamefont
  {Ando}}]{Wang2023}%
  \BibitemOpen
  \bibfield  {author} {\bibinfo {author} {\bibfnamefont {Y.}~\bibnamefont
  {Wang}}, \bibinfo {author} {\bibfnamefont {T.}~\bibnamefont {B{\"o}merich}},
  \bibinfo {author} {\bibfnamefont {J.}~\bibnamefont {Park}}, \bibinfo {author}
  {\bibfnamefont {H.~F.}\ \bibnamefont {Legg}}, \bibinfo {author}
  {\bibfnamefont {A.}~\bibnamefont {Taskin}}, \bibinfo {author} {\bibfnamefont
  {A.}~\bibnamefont {Rosch}},\ and\ \bibinfo {author} {\bibfnamefont
  {Y.}~\bibnamefont {Ando}},\ }\bibfield  {title} {\bibinfo {title} {Nonlinear
  {T}ransport due to {M}agnetic-{F}ield-{I}nduced {F}lat {B}ands in the
  {N}odal-{L}ine {S}emimetal {ZrTe$_5$}},\ }\href@noop {} {\bibfield  {journal}
  {\bibinfo  {journal} {Phys. Rev. Lett.}\ }\textbf {\bibinfo {volume} {131}},\
  \bibinfo {pages} {146602} (\bibinfo {year} {2023})}\BibitemShut {NoStop}%
\bibitem [{\citenamefont {Wang}\ \emph {et~al.}(2024)\citenamefont {Wang},
  \citenamefont {Huang}, \citenamefont {Liu}, \citenamefont {Feng},
  \citenamefont {Zhu}, \citenamefont {Wu}, \citenamefont {Xiao},\ and\
  \citenamefont {Yang}}]{Wang2024}%
  \BibitemOpen
  \bibfield  {author} {\bibinfo {author} {\bibfnamefont {H.}~\bibnamefont
  {Wang}}, \bibinfo {author} {\bibfnamefont {Y.-X.}\ \bibnamefont {Huang}},
  \bibinfo {author} {\bibfnamefont {H.}~\bibnamefont {Liu}}, \bibinfo {author}
  {\bibfnamefont {X.}~\bibnamefont {Feng}}, \bibinfo {author} {\bibfnamefont
  {J.}~\bibnamefont {Zhu}}, \bibinfo {author} {\bibfnamefont {W.}~\bibnamefont
  {Wu}}, \bibinfo {author} {\bibfnamefont {C.}~\bibnamefont {Xiao}},\ and\
  \bibinfo {author} {\bibfnamefont {S.~A.}\ \bibnamefont {Yang}},\ }\bibfield
  {title} {\bibinfo {title} {Orbital {O}rigin of the {I}ntrinsic {P}lanar
  {H}all {E}ffect},\ }\href@noop {} {\bibfield  {journal} {\bibinfo  {journal}
  {Phys. Rev. Lett.}\ }\textbf {\bibinfo {volume} {132}},\ \bibinfo {pages}
  {056301} (\bibinfo {year} {2024})}\BibitemShut {NoStop}%
\bibitem [{\citenamefont {Kikugawa}\ \emph {et~al.}(2016)\citenamefont
  {Kikugawa}, \citenamefont {Goswami}, \citenamefont {Kiswandhi}, \citenamefont
  {Choi}, \citenamefont {Graf}, \citenamefont {Baumbach}, \citenamefont
  {Brooks}, \citenamefont {Sugii}, \citenamefont {Iida}, \citenamefont
  {Nishio}, \citenamefont {Uji}, \citenamefont {Terashima}, \citenamefont
  {Rourke}, \citenamefont {Hussey}, \citenamefont {Takatsu}, \citenamefont
  {Yonezawa}, \citenamefont {Maeno},\ and\ \citenamefont
  {Balicas}}]{Kikugawa2016}%
  \BibitemOpen
  \bibfield  {author} {\bibinfo {author} {\bibfnamefont {N.}~\bibnamefont
  {Kikugawa}}, \bibinfo {author} {\bibfnamefont {P.}~\bibnamefont {Goswami}},
  \bibinfo {author} {\bibfnamefont {A.}~\bibnamefont {Kiswandhi}}, \bibinfo
  {author} {\bibfnamefont {E.~S.}\ \bibnamefont {Choi}}, \bibinfo {author}
  {\bibfnamefont {D.}~\bibnamefont {Graf}}, \bibinfo {author} {\bibfnamefont
  {R.~E.}\ \bibnamefont {Baumbach}}, \bibinfo {author} {\bibfnamefont {J.~S.}\
  \bibnamefont {Brooks}}, \bibinfo {author} {\bibfnamefont {K.}~\bibnamefont
  {Sugii}}, \bibinfo {author} {\bibfnamefont {Y.}~\bibnamefont {Iida}},
  \bibinfo {author} {\bibfnamefont {M.}~\bibnamefont {Nishio}}, \bibinfo
  {author} {\bibfnamefont {S.}~\bibnamefont {Uji}}, \bibinfo {author}
  {\bibfnamefont {T.}~\bibnamefont {Terashima}}, \bibinfo {author}
  {\bibfnamefont {P.~M.~C.}\ \bibnamefont {Rourke}}, \bibinfo {author}
  {\bibfnamefont {N.~E.}\ \bibnamefont {Hussey}}, \bibinfo {author}
  {\bibfnamefont {H.}~\bibnamefont {Takatsu}}, \bibinfo {author} {\bibfnamefont
  {S.}~\bibnamefont {Yonezawa}}, \bibinfo {author} {\bibfnamefont
  {Y.}~\bibnamefont {Maeno}},\ and\ \bibinfo {author} {\bibfnamefont
  {L.}~\bibnamefont {Balicas}},\ }\bibfield  {title} {\bibinfo {title}
  {Interplanar coupling-dependent magnetoresistivity in high-purity layered
  metals},\ }\href {https://doi.org/10.1038/ncomms10903} {\bibfield  {journal}
  {\bibinfo  {journal} {Nat. Commun.}\ }\textbf {\bibinfo {volume} {7}},\
  \bibinfo {pages} {10903} (\bibinfo {year} {2016})}\BibitemShut {NoStop}%
\bibitem [{\citenamefont {Li}\ \emph {et~al.}(2018{\natexlab{a}})\citenamefont
  {Li}, \citenamefont {Guo}, \citenamefont {Fu}, \citenamefont {Pan},
  \citenamefont {Wang}, \citenamefont {Ran}, \citenamefont {Bao}, \citenamefont
  {Ma}, \citenamefont {Cai}, \citenamefont {Wang}, \citenamefont {Yu},
  \citenamefont {Sun}, \citenamefont {Song},\ and\ \citenamefont
  {Wen}}]{Li2018b}%
  \BibitemOpen
  \bibfield  {author} {\bibinfo {author} {\bibfnamefont {S.}~\bibnamefont
  {Li}}, \bibinfo {author} {\bibfnamefont {Z.}~\bibnamefont {Guo}}, \bibinfo
  {author} {\bibfnamefont {D.}~\bibnamefont {Fu}}, \bibinfo {author}
  {\bibfnamefont {X.-C.}\ \bibnamefont {Pan}}, \bibinfo {author} {\bibfnamefont
  {J.}~\bibnamefont {Wang}}, \bibinfo {author} {\bibfnamefont {K.}~\bibnamefont
  {Ran}}, \bibinfo {author} {\bibfnamefont {S.}~\bibnamefont {Bao}}, \bibinfo
  {author} {\bibfnamefont {Z.}~\bibnamefont {Ma}}, \bibinfo {author}
  {\bibfnamefont {Z.}~\bibnamefont {Cai}}, \bibinfo {author} {\bibfnamefont
  {R.}~\bibnamefont {Wang}}, \bibinfo {author} {\bibfnamefont {R.}~\bibnamefont
  {Yu}}, \bibinfo {author} {\bibfnamefont {J.}~\bibnamefont {Sun}}, \bibinfo
  {author} {\bibfnamefont {F.}~\bibnamefont {Song}},\ and\ \bibinfo {author}
  {\bibfnamefont {J.}~\bibnamefont {Wen}},\ }\bibfield  {title} {\bibinfo
  {title} {Evidence for a {D}irac nodal-line semimetal in {SrAs$_3$}},\ }\href
  {https://doi.org/https://doi.org/10.1016/j.scib.2018.04.011} {\bibfield
  {journal} {\bibinfo  {journal} {Sci. Bull.}\ }\textbf {\bibinfo {volume}
  {63}},\ \bibinfo {pages} {535} (\bibinfo {year}
  {2018}{\natexlab{a}})}\BibitemShut {NoStop}%
\bibitem [{\citenamefont {Kokado}\ \emph {et~al.}(2012)\citenamefont {Kokado},
  \citenamefont {Tsunoda}, \citenamefont {Harigaya},\ and\ \citenamefont
  {Sakuma}}]{Kokado2012}%
  \BibitemOpen
  \bibfield  {author} {\bibinfo {author} {\bibfnamefont {S.}~\bibnamefont
  {Kokado}}, \bibinfo {author} {\bibfnamefont {M.}~\bibnamefont {Tsunoda}},
  \bibinfo {author} {\bibfnamefont {K.}~\bibnamefont {Harigaya}},\ and\
  \bibinfo {author} {\bibfnamefont {A.}~\bibnamefont {Sakuma}},\ }\bibfield
  {title} {\bibinfo {title} {Anisotropic {M}agnetoresistance {E}ffects in {F}e,
  {C}o, {N}i, {F}e$_4${N}, and {H}alf-{M}etallic {F}erromagnet: {A}
  {S}ystematic {A}nalysis},\ }\href@noop {} {\bibfield  {journal} {\bibinfo
  {journal} {J. Phys. Soc. Jpn.}\ }\textbf {\bibinfo {volume} {81}},\ \bibinfo
  {pages} {024705} (\bibinfo {year} {2012})}\BibitemShut {NoStop}%
\bibitem [{\citenamefont {Taskin}\ \emph {et~al.}(2017)\citenamefont {Taskin},
  \citenamefont {Legg}, \citenamefont {Yang}, \citenamefont {Sasaki},
  \citenamefont {Kanai}, \citenamefont {Matsumoto}, \citenamefont {Rosch},\
  and\ \citenamefont {Ando}}]{Taskin2017}%
  \BibitemOpen
  \bibfield  {author} {\bibinfo {author} {\bibfnamefont {A.}~\bibnamefont
  {Taskin}}, \bibinfo {author} {\bibfnamefont {H.~F.}\ \bibnamefont {Legg}},
  \bibinfo {author} {\bibfnamefont {F.}~\bibnamefont {Yang}}, \bibinfo {author}
  {\bibfnamefont {S.}~\bibnamefont {Sasaki}}, \bibinfo {author} {\bibfnamefont
  {Y.}~\bibnamefont {Kanai}}, \bibinfo {author} {\bibfnamefont
  {K.}~\bibnamefont {Matsumoto}}, \bibinfo {author} {\bibfnamefont
  {A.}~\bibnamefont {Rosch}},\ and\ \bibinfo {author} {\bibfnamefont
  {Y.}~\bibnamefont {Ando}},\ }\bibfield  {title} {\bibinfo {title} {Planar
  {H}all effect from the surface of topological insulators},\ }\href@noop {}
  {\bibfield  {journal} {\bibinfo  {journal} {Nat. Commun.}\ }\textbf {\bibinfo
  {volume} {8}},\ \bibinfo {pages} {1340} (\bibinfo {year} {2017})}\BibitemShut
  {NoStop}%
\bibitem [{\citenamefont {Li}\ \emph {et~al.}(2018{\natexlab{b}})\citenamefont
  {Li}, \citenamefont {Zhang}, \citenamefont {Zhang}, \citenamefont {Wen},\
  and\ \citenamefont {Zhang}}]{Li2018}%
  \BibitemOpen
  \bibfield  {author} {\bibinfo {author} {\bibfnamefont {P.}~\bibnamefont
  {Li}}, \bibinfo {author} {\bibfnamefont {C.}~\bibnamefont {Zhang}}, \bibinfo
  {author} {\bibfnamefont {J.}~\bibnamefont {Zhang}}, \bibinfo {author}
  {\bibfnamefont {Y.}~\bibnamefont {Wen}},\ and\ \bibinfo {author}
  {\bibfnamefont {X.}~\bibnamefont {Zhang}},\ }\bibfield  {title} {\bibinfo
  {title} {Giant planar {H}all effect in the {D}irac semimetal
  {ZrTe$_{5-\delta}$}},\ }\href@noop {} {\bibfield  {journal} {\bibinfo
  {journal} {Phys. Rev. B}\ }\textbf {\bibinfo {volume} {98}},\ \bibinfo
  {pages} {121108} (\bibinfo {year} {2018}{\natexlab{b}})}\BibitemShut
  {NoStop}%
\bibitem [{\citenamefont {Kumar}\ \emph {et~al.}(2018)\citenamefont {Kumar},
  \citenamefont {Guin}, \citenamefont {Felser},\ and\ \citenamefont
  {Shekhar}}]{Kumar2018}%
  \BibitemOpen
  \bibfield  {author} {\bibinfo {author} {\bibfnamefont {N.}~\bibnamefont
  {Kumar}}, \bibinfo {author} {\bibfnamefont {S.~N.}\ \bibnamefont {Guin}},
  \bibinfo {author} {\bibfnamefont {C.}~\bibnamefont {Felser}},\ and\ \bibinfo
  {author} {\bibfnamefont {C.}~\bibnamefont {Shekhar}},\ }\bibfield  {title}
  {\bibinfo {title} {Planar {H}all effect in the {W}eyl semimetal {GdPtBi}},\
  }\href@noop {} {\bibfield  {journal} {\bibinfo  {journal} {Phys. Rev. B}\
  }\textbf {\bibinfo {volume} {98}},\ \bibinfo {pages} {041103} (\bibinfo
  {year} {2018})}\BibitemShut {NoStop}%
\bibitem [{\citenamefont {Li}\ \emph {et~al.}(2018{\natexlab{c}})\citenamefont
  {Li}, \citenamefont {Wang}, \citenamefont {He}, \citenamefont {Wang},\ and\
  \citenamefont {Shen}}]{Li2018a}%
  \BibitemOpen
  \bibfield  {author} {\bibinfo {author} {\bibfnamefont {H.}~\bibnamefont
  {Li}}, \bibinfo {author} {\bibfnamefont {H.-W.}\ \bibnamefont {Wang}},
  \bibinfo {author} {\bibfnamefont {H.}~\bibnamefont {He}}, \bibinfo {author}
  {\bibfnamefont {J.}~\bibnamefont {Wang}},\ and\ \bibinfo {author}
  {\bibfnamefont {S.-Q.}\ \bibnamefont {Shen}},\ }\bibfield  {title} {\bibinfo
  {title} {Giant anisotropic magnetoresistance and planar {H}all effect in the
  {D}irac semimetal {Cd$_3$As$_2$}},\ }\href@noop {} {\bibfield  {journal}
  {\bibinfo  {journal} {Phys. Rev. B}\ }\textbf {\bibinfo {volume} {97}},\
  \bibinfo {pages} {201110} (\bibinfo {year} {2018}{\natexlab{c}})}\BibitemShut
  {NoStop}%
\bibitem [{\citenamefont {Wu}\ \emph {et~al.}(2018)\citenamefont {Wu},
  \citenamefont {Zheng}, \citenamefont {Chu}, \citenamefont {Liu},
  \citenamefont {Gao}, \citenamefont {Zhang}, \citenamefont {Lu}, \citenamefont
  {Han}, \citenamefont {Zhou}, \citenamefont {Ning} \emph {et~al.}}]{Wu2018}%
  \BibitemOpen
  \bibfield  {author} {\bibinfo {author} {\bibfnamefont {M.}~\bibnamefont
  {Wu}}, \bibinfo {author} {\bibfnamefont {G.}~\bibnamefont {Zheng}}, \bibinfo
  {author} {\bibfnamefont {W.}~\bibnamefont {Chu}}, \bibinfo {author}
  {\bibfnamefont {Y.}~\bibnamefont {Liu}}, \bibinfo {author} {\bibfnamefont
  {W.}~\bibnamefont {Gao}}, \bibinfo {author} {\bibfnamefont {H.}~\bibnamefont
  {Zhang}}, \bibinfo {author} {\bibfnamefont {J.}~\bibnamefont {Lu}}, \bibinfo
  {author} {\bibfnamefont {Y.}~\bibnamefont {Han}}, \bibinfo {author}
  {\bibfnamefont {J.}~\bibnamefont {Zhou}}, \bibinfo {author} {\bibfnamefont
  {W.}~\bibnamefont {Ning}}, \emph {et~al.},\ }\bibfield  {title} {\bibinfo
  {title} {Probing the chiral anomaly by planar {H}all effect in {D}irac
  semimetal {Cd$_3$As$_2$} nanoplates},\ }\href@noop {} {\bibfield  {journal}
  {\bibinfo  {journal} {Phys. Rev. B}\ }\textbf {\bibinfo {volume} {98}},\
  \bibinfo {pages} {161110} (\bibinfo {year} {2018})}\BibitemShut {NoStop}%
\bibitem [{\citenamefont {Chen}\ \emph {et~al.}(2018)\citenamefont {Chen},
  \citenamefont {Luo}, \citenamefont {Yan}, \citenamefont {Sun}, \citenamefont
  {Lv}, \citenamefont {Lu}, \citenamefont {Xi}, \citenamefont {Tong},
  \citenamefont {Sheng}, \citenamefont {Zhu} \emph {et~al.}}]{Chen2018}%
  \BibitemOpen
  \bibfield  {author} {\bibinfo {author} {\bibfnamefont {F.}~\bibnamefont
  {Chen}}, \bibinfo {author} {\bibfnamefont {X.}~\bibnamefont {Luo}}, \bibinfo
  {author} {\bibfnamefont {J.}~\bibnamefont {Yan}}, \bibinfo {author}
  {\bibfnamefont {Y.}~\bibnamefont {Sun}}, \bibinfo {author} {\bibfnamefont
  {H.}~\bibnamefont {Lv}}, \bibinfo {author} {\bibfnamefont {W.}~\bibnamefont
  {Lu}}, \bibinfo {author} {\bibfnamefont {C.}~\bibnamefont {Xi}}, \bibinfo
  {author} {\bibfnamefont {P.}~\bibnamefont {Tong}}, \bibinfo {author}
  {\bibfnamefont {Z.}~\bibnamefont {Sheng}}, \bibinfo {author} {\bibfnamefont
  {X.}~\bibnamefont {Zhu}}, \emph {et~al.},\ }\bibfield  {title} {\bibinfo
  {title} {Planar {H}all effect in the type-{II} {W}eyl semimetal
  {$T_d$-MoTe$_2$}},\ }\href@noop {} {\bibfield  {journal} {\bibinfo  {journal}
  {Phys. Rev. B}\ }\textbf {\bibinfo {volume} {98}},\ \bibinfo {pages} {041114}
  (\bibinfo {year} {2018})}\BibitemShut {NoStop}%
\bibitem [{\citenamefont {Liang}\ \emph {et~al.}(2019)\citenamefont {Liang},
  \citenamefont {Wang}, \citenamefont {Zhen}, \citenamefont {Yang},
  \citenamefont {Weng}, \citenamefont {Yan}, \citenamefont {Han}, \citenamefont
  {Tong}, \citenamefont {Zhu}, \citenamefont {Pi} \emph {et~al.}}]{Liang2019}%
  \BibitemOpen
  \bibfield  {author} {\bibinfo {author} {\bibfnamefont {D.}~\bibnamefont
  {Liang}}, \bibinfo {author} {\bibfnamefont {Y.}~\bibnamefont {Wang}},
  \bibinfo {author} {\bibfnamefont {W.}~\bibnamefont {Zhen}}, \bibinfo {author}
  {\bibfnamefont {J.}~\bibnamefont {Yang}}, \bibinfo {author} {\bibfnamefont
  {S.}~\bibnamefont {Weng}}, \bibinfo {author} {\bibfnamefont {X.}~\bibnamefont
  {Yan}}, \bibinfo {author} {\bibfnamefont {Y.}~\bibnamefont {Han}}, \bibinfo
  {author} {\bibfnamefont {W.}~\bibnamefont {Tong}}, \bibinfo {author}
  {\bibfnamefont {W.}~\bibnamefont {Zhu}}, \bibinfo {author} {\bibfnamefont
  {L.}~\bibnamefont {Pi}}, \emph {et~al.},\ }\bibfield  {title} {\bibinfo
  {title} {Origin of planar {H}all effect in type-{II} {W}eyl semimetal
  {M}o{T}e$_2$},\ }\href@noop {} {\bibfield  {journal} {\bibinfo  {journal}
  {Aip Adv.}\ }\textbf {\bibinfo {volume} {9}} (\bibinfo {year}
  {2019})}\BibitemShut {NoStop}%
\bibitem [{\citenamefont {Li}\ \emph {et~al.}(2019)\citenamefont {Li},
  \citenamefont {Zhang}, \citenamefont {Wen}, \citenamefont {Cheng},
  \citenamefont {Nichols}, \citenamefont {Cory}, \citenamefont {Miao},\ and\
  \citenamefont {Zhang}}]{Li2019}%
  \BibitemOpen
  \bibfield  {author} {\bibinfo {author} {\bibfnamefont {P.}~\bibnamefont
  {Li}}, \bibinfo {author} {\bibfnamefont {C.}~\bibnamefont {Zhang}}, \bibinfo
  {author} {\bibfnamefont {Y.}~\bibnamefont {Wen}}, \bibinfo {author}
  {\bibfnamefont {L.}~\bibnamefont {Cheng}}, \bibinfo {author} {\bibfnamefont
  {G.}~\bibnamefont {Nichols}}, \bibinfo {author} {\bibfnamefont {D.~G.}\
  \bibnamefont {Cory}}, \bibinfo {author} {\bibfnamefont {G.-X.}\ \bibnamefont
  {Miao}},\ and\ \bibinfo {author} {\bibfnamefont {X.-X.}\ \bibnamefont
  {Zhang}},\ }\bibfield  {title} {\bibinfo {title} {Anisotropic planar {H}all
  effect in the type-{II} topological {W}eyl semimetal {WTe$_2$}},\ }\href@noop
  {} {\bibfield  {journal} {\bibinfo  {journal} {Phys. Rev. B}\ }\textbf
  {\bibinfo {volume} {100}},\ \bibinfo {pages} {205128} (\bibinfo {year}
  {2019})}\BibitemShut {NoStop}%
\bibitem [{\citenamefont {Yang}\ \emph {et~al.}(2020)\citenamefont {Yang},
  \citenamefont {Noky}, \citenamefont {Gayles}, \citenamefont {Dejene},
  \citenamefont {Sun}, \citenamefont {D{\"o}rr}, \citenamefont {Skourski},
  \citenamefont {Felser}, \citenamefont {Ali}, \citenamefont {Liu} \emph
  {et~al.}}]{Yang2020}%
  \BibitemOpen
  \bibfield  {author} {\bibinfo {author} {\bibfnamefont {S.-Y.}\ \bibnamefont
  {Yang}}, \bibinfo {author} {\bibfnamefont {J.}~\bibnamefont {Noky}}, \bibinfo
  {author} {\bibfnamefont {J.}~\bibnamefont {Gayles}}, \bibinfo {author}
  {\bibfnamefont {F.~K.}\ \bibnamefont {Dejene}}, \bibinfo {author}
  {\bibfnamefont {Y.}~\bibnamefont {Sun}}, \bibinfo {author} {\bibfnamefont
  {M.}~\bibnamefont {D{\"o}rr}}, \bibinfo {author} {\bibfnamefont
  {Y.}~\bibnamefont {Skourski}}, \bibinfo {author} {\bibfnamefont
  {C.}~\bibnamefont {Felser}}, \bibinfo {author} {\bibfnamefont {M.~N.}\
  \bibnamefont {Ali}}, \bibinfo {author} {\bibfnamefont {E.}~\bibnamefont
  {Liu}}, \emph {et~al.},\ }\bibfield  {title} {\bibinfo {title}
  {Field-modulated anomalous {H}all conductivity and planar {H}all effect in
  {Co$_3$Sn$_2$S$_2$} nanoflakes},\ }\href@noop {} {\bibfield  {journal}
  {\bibinfo  {journal} {Nano Lett.}\ }\textbf {\bibinfo {volume} {20}},\
  \bibinfo {pages} {7860} (\bibinfo {year} {2020})}\BibitemShut {NoStop}%
\bibitem [{SM()}]{SM}%
  \BibitemOpen
  \href@noop {} {\bibinfo  {journal} {See {S}upplemental {M}aterial for
  additional data and discussion}\ }\BibitemShut {NoStop}%
\bibitem [{\citenamefont {Xu}\ \emph {et~al.}(2018)\citenamefont {Xu},
  \citenamefont {Zhao}, \citenamefont {Marsik}, \citenamefont {Sheveleva},
  \citenamefont {Lyzwa}, \citenamefont {Dai}, \citenamefont {Chen},
  \citenamefont {Qiu},\ and\ \citenamefont {Bernhard}}]{Xu2018}%
  \BibitemOpen
\bibfield  {journal} {  }\bibfield  {author} {\bibinfo {author} {\bibfnamefont
  {B.}~\bibnamefont {Xu}}, \bibinfo {author} {\bibfnamefont {L.~X.}\
  \bibnamefont {Zhao}}, \bibinfo {author} {\bibfnamefont {P.}~\bibnamefont
  {Marsik}}, \bibinfo {author} {\bibfnamefont {E.}~\bibnamefont {Sheveleva}},
  \bibinfo {author} {\bibfnamefont {F.}~\bibnamefont {Lyzwa}}, \bibinfo
  {author} {\bibfnamefont {Y.~M.}\ \bibnamefont {Dai}}, \bibinfo {author}
  {\bibfnamefont {G.~F.}\ \bibnamefont {Chen}}, \bibinfo {author}
  {\bibfnamefont {X.~G.}\ \bibnamefont {Qiu}},\ and\ \bibinfo {author}
  {\bibfnamefont {C.}~\bibnamefont {Bernhard}},\ }\bibfield  {title} {\bibinfo
  {title} {Temperature-{D}riven {T}opological {P}hase {T}ransition and
  {I}ntermediate {D}irac {S}emimetal {P}hase in {Z}r{T}e$_5$},\ }\href
  {https://doi.org/10.1103/PhysRevLett.121.187401} {\bibfield  {journal}
  {\bibinfo  {journal} {Phys. Rev. Lett.}\ }\textbf {\bibinfo {volume} {121}},\
  \bibinfo {pages} {187401} (\bibinfo {year} {2018})}\BibitemShut {NoStop}%
\bibitem [{\citenamefont {Liang}\ \emph {et~al.}(2018)\citenamefont {Liang},
  \citenamefont {Lin}, \citenamefont {Gibson}, \citenamefont {Kushwaha},
  \citenamefont {Liu}, \citenamefont {Wang}, \citenamefont {Xiong},
  \citenamefont {Sobota}, \citenamefont {Hashimoto}, \citenamefont {Kirchmann},
  \citenamefont {Shen}, \citenamefont {Cava},\ and\ \citenamefont
  {Ong}}]{Liang2018}%
  \BibitemOpen
  \bibfield  {author} {\bibinfo {author} {\bibfnamefont {T.}~\bibnamefont
  {Liang}}, \bibinfo {author} {\bibfnamefont {J.}~\bibnamefont {Lin}}, \bibinfo
  {author} {\bibfnamefont {Q.}~\bibnamefont {Gibson}}, \bibinfo {author}
  {\bibfnamefont {S.}~\bibnamefont {Kushwaha}}, \bibinfo {author}
  {\bibfnamefont {M.}~\bibnamefont {Liu}}, \bibinfo {author} {\bibfnamefont
  {W.}~\bibnamefont {Wang}}, \bibinfo {author} {\bibfnamefont {H.}~\bibnamefont
  {Xiong}}, \bibinfo {author} {\bibfnamefont {J.~A.}\ \bibnamefont {Sobota}},
  \bibinfo {author} {\bibfnamefont {M.}~\bibnamefont {Hashimoto}}, \bibinfo
  {author} {\bibfnamefont {P.~S.}\ \bibnamefont {Kirchmann}}, \bibinfo {author}
  {\bibfnamefont {Z.-X.}\ \bibnamefont {Shen}}, \bibinfo {author}
  {\bibfnamefont {R.~J.}\ \bibnamefont {Cava}},\ and\ \bibinfo {author}
  {\bibfnamefont {N.~P.}\ \bibnamefont {Ong}},\ }\bibfield  {title} {\bibinfo
  {title} {Anomalous {H}all effect in {Z}r{T}e$_5$},\ }\href
  {https://doi.org/10.1038/s41567-018-0078-z} {\bibfield  {journal} {\bibinfo
  {journal} {Nat. Phys.}\ }\textbf {\bibinfo {volume} {14}},\ \bibinfo {pages}
  {451} (\bibinfo {year} {2018})}\BibitemShut {NoStop}%
\bibitem [{\citenamefont {Tang}\ \emph {et~al.}(2019)\citenamefont {Tang},
  \citenamefont {Ren}, \citenamefont {Wang}, \citenamefont {Zhong},
  \citenamefont {Schneeloch}, \citenamefont {Yang}, \citenamefont {Yang},
  \citenamefont {Lee}, \citenamefont {Gu}, \citenamefont {Qiao},\ and\
  \citenamefont {Zhang}}]{Tang2019}%
  \BibitemOpen
  \bibfield  {author} {\bibinfo {author} {\bibfnamefont {F.}~\bibnamefont
  {Tang}}, \bibinfo {author} {\bibfnamefont {Y.}~\bibnamefont {Ren}}, \bibinfo
  {author} {\bibfnamefont {P.}~\bibnamefont {Wang}}, \bibinfo {author}
  {\bibfnamefont {R.}~\bibnamefont {Zhong}}, \bibinfo {author} {\bibfnamefont
  {J.}~\bibnamefont {Schneeloch}}, \bibinfo {author} {\bibfnamefont {S.~A.}\
  \bibnamefont {Yang}}, \bibinfo {author} {\bibfnamefont {K.}~\bibnamefont
  {Yang}}, \bibinfo {author} {\bibfnamefont {P.~A.}\ \bibnamefont {Lee}},
  \bibinfo {author} {\bibfnamefont {G.}~\bibnamefont {Gu}}, \bibinfo {author}
  {\bibfnamefont {Z.}~\bibnamefont {Qiao}},\ and\ \bibinfo {author}
  {\bibfnamefont {L.}~\bibnamefont {Zhang}},\ }\bibfield  {title} {\bibinfo
  {title} {Three-dimensional quantum {H}all effect and metal-insulator
  transition in {Z}r{T}e$_5$},\ }\href
  {https://doi.org/10.1038/s41586-019-1180-9} {\bibfield  {journal} {\bibinfo
  {journal} {Nature}\ }\textbf {\bibinfo {volume} {569}},\ \bibinfo {pages}
  {537} (\bibinfo {year} {2019})}\BibitemShut {NoStop}%
\bibitem [{\citenamefont {Jiang}\ \emph {et~al.}(2020)\citenamefont {Jiang},
  \citenamefont {Wang}, \citenamefont {Zhao}, \citenamefont {Dun},
  \citenamefont {Huang}, \citenamefont {Wu}, \citenamefont {Mourigal},
  \citenamefont {Zhou}, \citenamefont {Pan}, \citenamefont {Ozerov} \emph
  {et~al.}}]{Jiang2020}%
  \BibitemOpen
  \bibfield  {author} {\bibinfo {author} {\bibfnamefont {Y.}~\bibnamefont
  {Jiang}}, \bibinfo {author} {\bibfnamefont {J.}~\bibnamefont {Wang}},
  \bibinfo {author} {\bibfnamefont {T.}~\bibnamefont {Zhao}}, \bibinfo {author}
  {\bibfnamefont {Z.}~\bibnamefont {Dun}}, \bibinfo {author} {\bibfnamefont
  {Q.}~\bibnamefont {Huang}}, \bibinfo {author} {\bibfnamefont
  {X.}~\bibnamefont {Wu}}, \bibinfo {author} {\bibfnamefont {M.}~\bibnamefont
  {Mourigal}}, \bibinfo {author} {\bibfnamefont {H.}~\bibnamefont {Zhou}},
  \bibinfo {author} {\bibfnamefont {W.}~\bibnamefont {Pan}}, \bibinfo {author}
  {\bibfnamefont {M.}~\bibnamefont {Ozerov}}, \emph {et~al.},\ }\bibfield
  {title} {\bibinfo {title} {Unraveling the topological phase of {ZrTe$_5$} via
  magnetoinfrared spectroscopy},\ }\href@noop {} {\bibfield  {journal}
  {\bibinfo  {journal} {Phys. Rev. Lett.}\ }\textbf {\bibinfo {volume} {125}},\
  \bibinfo {pages} {046403} (\bibinfo {year} {2020})}\BibitemShut {NoStop}%
\bibitem [{\citenamefont {Levin}(1997)}]{Levin1997}%
  \BibitemOpen
  \bibfield  {author} {\bibinfo {author} {\bibfnamefont {G.}~\bibnamefont
  {Levin}},\ }\bibfield  {title} {\bibinfo {title} {On the theory of
  measurement of anisotropic electrical resistivity by flux transformer
  method},\ }\href@noop {} {\bibfield  {journal} {\bibinfo  {journal} {J. Appl.
  Phys.}\ }\textbf {\bibinfo {volume} {81}},\ \bibinfo {pages} {714} (\bibinfo
  {year} {1997})}\BibitemShut {NoStop}%
\bibitem [{\citenamefont {Busch}\ \emph {et~al.}(1992)\citenamefont {Busch},
  \citenamefont {Ries}, \citenamefont {Werthner}, \citenamefont
  {Kreiselmeyer},\ and\ \citenamefont {Saemann-Ischenko}}]{Busch1992}%
  \BibitemOpen
  \bibfield  {author} {\bibinfo {author} {\bibfnamefont {R.}~\bibnamefont
  {Busch}}, \bibinfo {author} {\bibfnamefont {G.}~\bibnamefont {Ries}},
  \bibinfo {author} {\bibfnamefont {H.}~\bibnamefont {Werthner}}, \bibinfo
  {author} {\bibfnamefont {G.}~\bibnamefont {Kreiselmeyer}},\ and\ \bibinfo
  {author} {\bibfnamefont {G.}~\bibnamefont {Saemann-Ischenko}},\ }\bibfield
  {title} {\bibinfo {title} {New aspects of the mixed state from six-terminal
  measurements on {Bi$_2$Sr$_2$CaCu$_2$O$_x$} single crystals},\ }\href@noop {}
  {\bibfield  {journal} {\bibinfo  {journal} {Phys. Rev. Lett.}\ }\textbf
  {\bibinfo {volume} {69}},\ \bibinfo {pages} {522} (\bibinfo {year}
  {1992})}\BibitemShut {NoStop}%
\bibitem [{\citenamefont {Tang}\ \emph {et~al.}(2003)\citenamefont {Tang},
  \citenamefont {Kawakami}, \citenamefont {Awschalom},\ and\ \citenamefont
  {Roukes}}]{Tang2003}%
  \BibitemOpen
  \bibfield  {author} {\bibinfo {author} {\bibfnamefont {H.}~\bibnamefont
  {Tang}}, \bibinfo {author} {\bibfnamefont {R.}~\bibnamefont {Kawakami}},
  \bibinfo {author} {\bibfnamefont {D.}~\bibnamefont {Awschalom}},\ and\
  \bibinfo {author} {\bibfnamefont {M.}~\bibnamefont {Roukes}},\ }\bibfield
  {title} {\bibinfo {title} {Giant {P}lanar {H}all {E}ffect in {E}pitaxial
  ({G}a, {M}n){A}s {D}evices},\ }\href@noop {} {\bibfield  {journal} {\bibinfo
  {journal} {Phys. Rev. Lett.}\ }\textbf {\bibinfo {volume} {90}},\ \bibinfo
  {pages} {107201} (\bibinfo {year} {2003})}\BibitemShut {NoStop}%
\bibitem [{\citenamefont {Yang}\ \emph {et~al.}(2019)\citenamefont {Yang},
  \citenamefont {Zhen}, \citenamefont {Liang}, \citenamefont {Wang},
  \citenamefont {Yan}, \citenamefont {Weng}, \citenamefont {Wang},
  \citenamefont {Tong}, \citenamefont {Pi}, \citenamefont {Zhu} \emph
  {et~al.}}]{Yang2019}%
  \BibitemOpen
  \bibfield  {author} {\bibinfo {author} {\bibfnamefont {J.}~\bibnamefont
  {Yang}}, \bibinfo {author} {\bibfnamefont {W.}~\bibnamefont {Zhen}}, \bibinfo
  {author} {\bibfnamefont {D.}~\bibnamefont {Liang}}, \bibinfo {author}
  {\bibfnamefont {Y.}~\bibnamefont {Wang}}, \bibinfo {author} {\bibfnamefont
  {X.}~\bibnamefont {Yan}}, \bibinfo {author} {\bibfnamefont {S.}~\bibnamefont
  {Weng}}, \bibinfo {author} {\bibfnamefont {J.}~\bibnamefont {Wang}}, \bibinfo
  {author} {\bibfnamefont {W.}~\bibnamefont {Tong}}, \bibinfo {author}
  {\bibfnamefont {L.}~\bibnamefont {Pi}}, \bibinfo {author} {\bibfnamefont
  {W.}~\bibnamefont {Zhu}}, \emph {et~al.},\ }\bibfield  {title} {\bibinfo
  {title} {Current jetting distorted planar {H}all effect in a {W}eyl semimetal
  with ultrahigh mobility},\ }\href@noop {} {\bibfield  {journal} {\bibinfo
  {journal} {Phys. Rev. Mater.}\ }\textbf {\bibinfo {volume} {3}},\ \bibinfo
  {pages} {014201} (\bibinfo {year} {2019})}\BibitemShut {NoStop}%
\bibitem [{\citenamefont {Wang}\ \emph {et~al.}(2025)\citenamefont {Wang},
  \citenamefont {Wowchik}, \citenamefont {B{\"o}merich}, \citenamefont
  {Taskin}, \citenamefont {Rosch},\ and\ \citenamefont {Ando}}]{Zenodo}%
  \BibitemOpen
  \bibfield  {author} {\bibinfo {author} {\bibfnamefont {Y.}~\bibnamefont
  {Wang}}, \bibinfo {author} {\bibfnamefont {A.}~\bibnamefont {Wowchik}},
  \bibinfo {author} {\bibfnamefont {T.}~\bibnamefont {B{\"o}merich}}, \bibinfo
  {author} {\bibfnamefont {A.}~\bibnamefont {Taskin}}, \bibinfo {author}
  {\bibfnamefont {A.}~\bibnamefont {Rosch}},\ and\ \bibinfo {author}
  {\bibfnamefont {Y.}~\bibnamefont {Ando}},\ }\bibfield  {title} {\bibinfo
  {title} {Generic chiral anomaly and planar {H}all effect in a non-{W}eyl
  system},\ }\href {https://doi.org/10.5281/zenodo.15634664}
  {10.5281/zenodo.15634664} (\bibinfo {year} {2025})\BibitemShut {NoStop}%
\end{thebibliography}%


\begin{thebibliography}{7}%
\makeatletter
\providecommand \@ifxundefined [1]{%
 \@ifx{#1\undefined}
}%
\providecommand \@ifnum [1]{%
 \ifnum #1\expandafter \@firstoftwo
 \else \expandafter \@secondoftwo
 \fi
}%
\providecommand \@ifx [1]{%
 \ifx #1\expandafter \@firstoftwo
 \else \expandafter \@secondoftwo
 \fi
}%
\providecommand \natexlab [1]{#1}%
\providecommand \enquote  [1]{``#1''}%
\providecommand \bibnamefont  [1]{#1}%
\providecommand \bibfnamefont [1]{#1}%
\providecommand \citenamefont [1]{#1}%
\providecommand \href@noop [0]{\@secondoftwo}%
\providecommand \href [0]{\begingroup \@sanitize@url \@href}%
\providecommand \@href[1]{\@@startlink{#1}\@@href}%
\providecommand \@@href[1]{\endgroup#1\@@endlink}%
\providecommand \@sanitize@url [0]{\catcode `\\12\catcode `\$12\catcode `\&12\catcode `\#12\catcode `\^12\catcode `\_12\catcode `\%12\relax}%
\providecommand \@@startlink[1]{}%
\providecommand \@@endlink[0]{}%
\providecommand \url  [0]{\begingroup\@sanitize@url \@url }%
\providecommand \@url [1]{\endgroup\@href {#1}{\urlprefix }}%
\providecommand \urlprefix  [0]{URL }%
\providecommand \Eprint [0]{\href }%
\providecommand \doibase [0]{https://doi.org/}%
\providecommand \selectlanguage [0]{\@gobble}%
\providecommand \bibinfo  [0]{\@secondoftwo}%
\providecommand \bibfield  [0]{\@secondoftwo}%
\providecommand \translation [1]{[#1]}%
\providecommand \BibitemOpen [0]{}%
\providecommand \bibitemStop [0]{}%
\providecommand \bibitemNoStop [0]{.\EOS\space}%
\providecommand \EOS [0]{\spacefactor3000\relax}%
\providecommand \BibitemShut  [1]{\csname bibitem#1\endcsname}%
\let\auto@bib@innerbib\@empty
%</preamble>
\bibitem [{\citenamefont {Wang}\ \emph {et~al.}(2022)\citenamefont {Wang}, \citenamefont {Legg}, \citenamefont {B{\"o}merich}, \citenamefont {Park}, \citenamefont {Biesenkamp}, \citenamefont {Taskin}, \citenamefont {Braden}, \citenamefont {Rosch},\ and\ \citenamefont {Ando}}]{Wang2022}%
  \BibitemOpen
  \bibfield  {author} {\bibinfo {author} {\bibfnamefont {Y.}~\bibnamefont {Wang}}, \bibinfo {author} {\bibfnamefont {H.~F.}\ \bibnamefont {Legg}}, \bibinfo {author} {\bibfnamefont {T.}~\bibnamefont {B{\"o}merich}}, \bibinfo {author} {\bibfnamefont {J.}~\bibnamefont {Park}}, \bibinfo {author} {\bibfnamefont {S.}~\bibnamefont {Biesenkamp}}, \bibinfo {author} {\bibfnamefont {A.}~\bibnamefont {Taskin}}, \bibinfo {author} {\bibfnamefont {M.}~\bibnamefont {Braden}}, \bibinfo {author} {\bibfnamefont {A.}~\bibnamefont {Rosch}},\ and\ \bibinfo {author} {\bibfnamefont {Y.}~\bibnamefont {Ando}},\ }\href@noop {} {\bibfield  {journal} {\bibinfo  {journal} {Phys. Rev. Lett.}\ }\textbf {\bibinfo {volume} {128}},\ \bibinfo {pages} {176602} (\bibinfo {year} {2022})}\BibitemShut {NoStop}%
\bibitem [{\citenamefont {Wang}\ \emph {et~al.}(2023)\citenamefont {Wang}, \citenamefont {B{\"o}merich}, \citenamefont {Park}, \citenamefont {Legg}, \citenamefont {Taskin}, \citenamefont {Rosch},\ and\ \citenamefont {Ando}}]{Wang2023}%
  \BibitemOpen
  \bibfield  {author} {\bibinfo {author} {\bibfnamefont {Y.}~\bibnamefont {Wang}}, \bibinfo {author} {\bibfnamefont {T.}~\bibnamefont {B{\"o}merich}}, \bibinfo {author} {\bibfnamefont {J.}~\bibnamefont {Park}}, \bibinfo {author} {\bibfnamefont {H.~F.}\ \bibnamefont {Legg}}, \bibinfo {author} {\bibfnamefont {A.}~\bibnamefont {Taskin}}, \bibinfo {author} {\bibfnamefont {A.}~\bibnamefont {Rosch}},\ and\ \bibinfo {author} {\bibfnamefont {Y.}~\bibnamefont {Ando}},\ }\href@noop {} {\bibfield  {journal} {\bibinfo  {journal} {Phys. Rev. Lett.}\ }\textbf {\bibinfo {volume} {131}},\ \bibinfo {pages} {146602} (\bibinfo {year} {2023})}\BibitemShut {NoStop}%
\bibitem [{\citenamefont {Wang}\ \emph {et~al.}(2025)\citenamefont {Wang}, \citenamefont {B{\"o}merich}, \citenamefont {Taskin}, \citenamefont {Rosch},\ and\ \citenamefont {Ando}}]{Wang2025}%
  \BibitemOpen
  \bibfield  {author} {\bibinfo {author} {\bibfnamefont {Y.}~\bibnamefont {Wang}}, \bibinfo {author} {\bibfnamefont {T.}~\bibnamefont {B{\"o}merich}}, \bibinfo {author} {\bibfnamefont {A.~A.}\ \bibnamefont {Taskin}}, \bibinfo {author} {\bibfnamefont {A.}~\bibnamefont {Rosch}},\ and\ \bibinfo {author} {\bibfnamefont {Y.}~\bibnamefont {Ando}},\ }\href {https://doi.org/10.1103/PhysRevB.111.L041201} {\bibfield  {journal} {\bibinfo  {journal} {Phys. Rev. B}\ }\textbf {\bibinfo {volume} {111}},\ \bibinfo {pages} {L041201} (\bibinfo {year} {2025})}\BibitemShut {NoStop}%
\bibitem [{\citenamefont {Burkov}(2017)}]{Burkov2017}%
  \BibitemOpen
  \bibfield  {author} {\bibinfo {author} {\bibfnamefont {A.}~\bibnamefont {Burkov}},\ }\href@noop {} {\bibfield  {journal} {\bibinfo  {journal} {Phys. Rev. B}\ }\textbf {\bibinfo {volume} {96}},\ \bibinfo {pages} {041110} (\bibinfo {year} {2017})}\BibitemShut {NoStop}%
\bibitem [{\citenamefont {Nandy}\ \emph {et~al.}(2017)\citenamefont {Nandy}, \citenamefont {Sharma}, \citenamefont {Taraphder},\ and\ \citenamefont {Tewari}}]{Nandy2017}%
  \BibitemOpen
  \bibfield  {author} {\bibinfo {author} {\bibfnamefont {S.}~\bibnamefont {Nandy}}, \bibinfo {author} {\bibfnamefont {G.}~\bibnamefont {Sharma}}, \bibinfo {author} {\bibfnamefont {A.}~\bibnamefont {Taraphder}},\ and\ \bibinfo {author} {\bibfnamefont {S.}~\bibnamefont {Tewari}},\ }\href@noop {} {\bibfield  {journal} {\bibinfo  {journal} {Phys. Rev. Lett.}\ }\textbf {\bibinfo {volume} {119}},\ \bibinfo {pages} {176804} (\bibinfo {year} {2017})}\BibitemShut {NoStop}%
\bibitem [{\citenamefont {Busch}\ \emph {et~al.}(1992)\citenamefont {Busch}, \citenamefont {Ries}, \citenamefont {Werthner}, \citenamefont {Kreiselmeyer},\ and\ \citenamefont {Saemann-Ischenko}}]{Busch1992}%
  \BibitemOpen
  \bibfield  {author} {\bibinfo {author} {\bibfnamefont {R.}~\bibnamefont {Busch}}, \bibinfo {author} {\bibfnamefont {G.}~\bibnamefont {Ries}}, \bibinfo {author} {\bibfnamefont {H.}~\bibnamefont {Werthner}}, \bibinfo {author} {\bibfnamefont {G.}~\bibnamefont {Kreiselmeyer}},\ and\ \bibinfo {author} {\bibfnamefont {G.}~\bibnamefont {Saemann-Ischenko}},\ }\href@noop {} {\bibfield  {journal} {\bibinfo  {journal} {Phys. Rev. Lett.}\ }\textbf {\bibinfo {volume} {69}},\ \bibinfo {pages} {522} (\bibinfo {year} {1992})}\BibitemShut {NoStop}%
\bibitem [{\citenamefont {Levin}(1997)}]{Levin1997}%
  \BibitemOpen
  \bibfield  {author} {\bibinfo {author} {\bibfnamefont {G.}~\bibnamefont {Levin}},\ }\href@noop {} {\bibfield  {journal} {\bibinfo  {journal} {J. Appl. Phys.}\ }\textbf {\bibinfo {volume} {81}},\ \bibinfo {pages} {714} (\bibinfo {year} {1997})}\BibitemShut {NoStop}%
\end{thebibliography}%

\end{document}

%{\bf{\large Supplemental Material for ``Unconventional Planar Hall Effect as a Clue to Understand Negative Longitudinal Magnetoresistance in ZrTe$_5$"}}

%\vspace{5mm}

\title{Supplemental Material for ``Generic Chiral Anomaly and Planar Hall Effect in a Non-Weyl System''}

\author{Yongjian Wang}
\affiliation{Physics Institute II, University of Cologne, D-50937 K\"oln, Germany}

\author{Alexander Wowchik}
\affiliation{Institute for Theoretical Physics, University of Cologne, Z\"ulpicher Str. 77, 50937 K\"oln, Germany}

\author{Thomas B\"omerich}
\affiliation{Institute for Theoretical Physics, University of Cologne, Z\"ulpicher Str. 77, 50937 K\"oln, Germany}

\author{A. A. Taskin}
\affiliation{Physics Institute II, University of Cologne, D-50937 K\"oln, Germany}

\author{Achim Rosch}
\affiliation{Institute for Theoretical Physics, University of Cologne, Z\"ulpicher Str. 77, 50937 K\"oln, Germany}

\author{Yoichi Ando}
\affiliation{Physics Institute II, University of Cologne, D-50937 K\"oln, Germany}

\maketitle

\section{Experimental details and additional data}

\subsection{Method}

Single crystals of ZrTe$_5$ with $T_{\rm p}$ of 0 K and 138 K were grown by a Te-flux method and a chemical vapor transport method, respectively. The details were reported in Ref. \cite{Wang2022}. To make good electrical contacts for the transport measurements, the surface of a bulk single crystal was cleaned by Ar plasma to remove the oxidized layer and Au/Nb contact electrodes were sputter-deposited, after which gold wires were bonded to the sample with silver paste. The voltage electrodes of samples A-E were prepared on the top and bottom of the naturally cleaved $ac$ surfaces of the ZrTe$_5$ samples. To protect the side surfaces of the sample from being short-circuited near the voltage contacts, a little GE varnish was used to cover the side surfaces near the voltage electrodes before the processes of in-situ cleaning, metal deposition, and gold-wire bonding. To make homogeneous current distribution in samples A and D during the measurements of the planar Hall effect (PHE), the $bc$ surfaces were covered by the current electrodes, which are very far away from the voltage electrodes (more than 500 $\mu$m). Samples B was prepared for the flux transformer method, for which the current contacts were made on the top surface of the naturally cleaved $ac$ surfaces (see Sec. \ref{sec:fluxTF} for details). Sample C was used for cross-checking of the magnetoresistance in $\rho_{bb}$ with a different contact configuration discussed in see Sec. \ref{sec:fluxTF}.

All the transport measurements were performed in a Quantum Design Physical Properties Measurement System (PPMS) with a rotating sample holder. The resistance and the Hall resistance were measured with a  low-frequency (17.33 Hz) AC lock-in technique.

\subsection{Additional data on the PHE in sample A}

\begin{figure}[h]
	\centering
	\includegraphics[width=14cm]{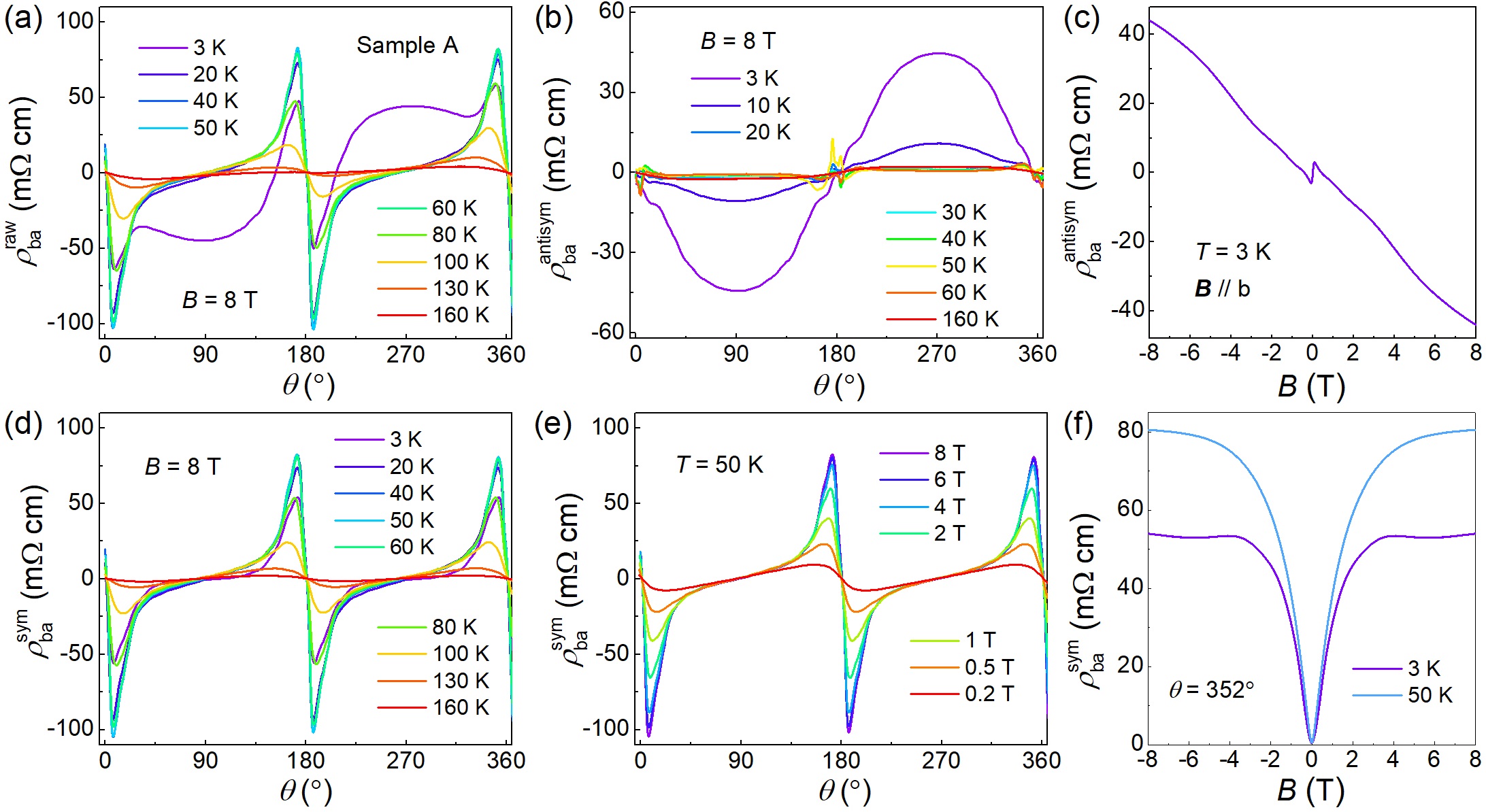}
	\caption{
	(a) $\theta$-dependence of the raw $\rho_{ba}$ data, $\rho_{ba}^{\rm raw}$, in 8 T. 
	(b) $\theta$-dependence of the antisymmetric component, $\rho_{ba}^{\rm antisym}$, in 8 T. 
	(c) $B$-dependence of $\rho_{ba}^{\rm antisym}$ for $\boldsymbol{B} \parallel b$-axis at 3 K. 
	(d) $\theta$-dependence of the symmetric component, $\rho_{ba}^{\rm sym}$, in 8 T. 
	(e) $\theta$-dependence of $\rho_{ba}^{\rm sym}$ for various strengths of $B$ at 50 K. 
	(f) $B$-dependence of $\rho_{ba}^{\rm sym}$ at 3 K and 50 K for $\theta = 352^\circ$, which is near the peak position.}
\end{figure}

The $\theta$-dependence of the raw $\rho_{ba}$ data is shown in Fig. S1(a) for various temperatures. We decomposed raw $\rho_{ba}$ into antisymmetric part $\rho_{ba}^{\rm antisym}$ [Fig. S1(b)] and symmetric part $\rho_{ba}^{\rm sym}$ [Fig. S1(d)]. The antisymmetric component can arise due to a misalignment of the magnetic-field rotation plane from the $ab$ plane. A random misalignment would lead to $\cos (\theta +\delta)$ dependence on the angle $\theta$ with a random phase shift of $\delta$. Interestingly, we found that $\rho_{ba}^{\rm antisym}$ shows a nearly-regular $\cos \theta$ dependence without a phase shift, suggesting an intrinsic origin. The $B$-dependence of $\rho_{ba}^{\rm antisym}$ is roughly linear [Fig. S1(c)]. This antisymmetric part is observed only below $\sim$10 K, which is similar to the magnetochiral anisotropy \cite{Wang2022} and nonlinear transport \cite{Wang2023} in ZrTe$_5$. It is possible that $\rho_{ba}^{\rm antisym}$ is due to Berry curvature, but it is also possible that $\rho_{ba}^{\rm antisym}$ is due to strong symmetry breaking as in the case of the recently-observed parallel-field Hall effect \cite{Wang2025}. 
More experimental efforts are necessary to elucidate the origin of $\rho_{ba}^{\rm antisym}$, which is beyond the scope in this paper. 

The amplitude of the PHE becomes maximum at $\sim$50 K [see Fig. S1(d)], where the largest negative LMR was also observed [see Fig. 2(d) of the main text]. With further increasing temperature, the sharp peak-and-dip feature weakens, and the PHE becomes almost undetectable at 160 K. Figure S1(e) shows how the sharp peak-and-dip feature at 50 K weakens as the magnetic field is reduced.
The magnetic-field dependence of the amplitude of the PHE, exemplified by the $\rho_{ba}^{\rm sym}$ value at $\theta$ = 352$^\circ$, is shown in Fig. S1(f) for 3 K and 50 K. The observed $B$-dependence is completely different from the $B^2$ behavior of the PHE in Weyl semimetals \cite{Burkov2017, Nandy2017}.

\subsection{Reproducibility of the negative longitudinal magnetoresistance in sample D}

The negative longitudinal magnetoresistance (LMR) was reproduced in sample D as shown in Fig. S2, which captures all the characteristic features of the negative LMR in ZrTe$_5$ as described in the main text. Compared to sample A, the main difference is that the negative LMR becomes most prominent at 10 K in sample D, which is possibly due to a variation in the carrier density between samples. 

%The amplitude of the PHE in sample D was maximum also at 10 K [Fig. S3(a)]. This correlation between negative LMR and PHE was observed in sample A as well (will be shown later), which suggests the same origin of them.

\begin{figure}[h]
	\centering
	\includegraphics[width=12cm]{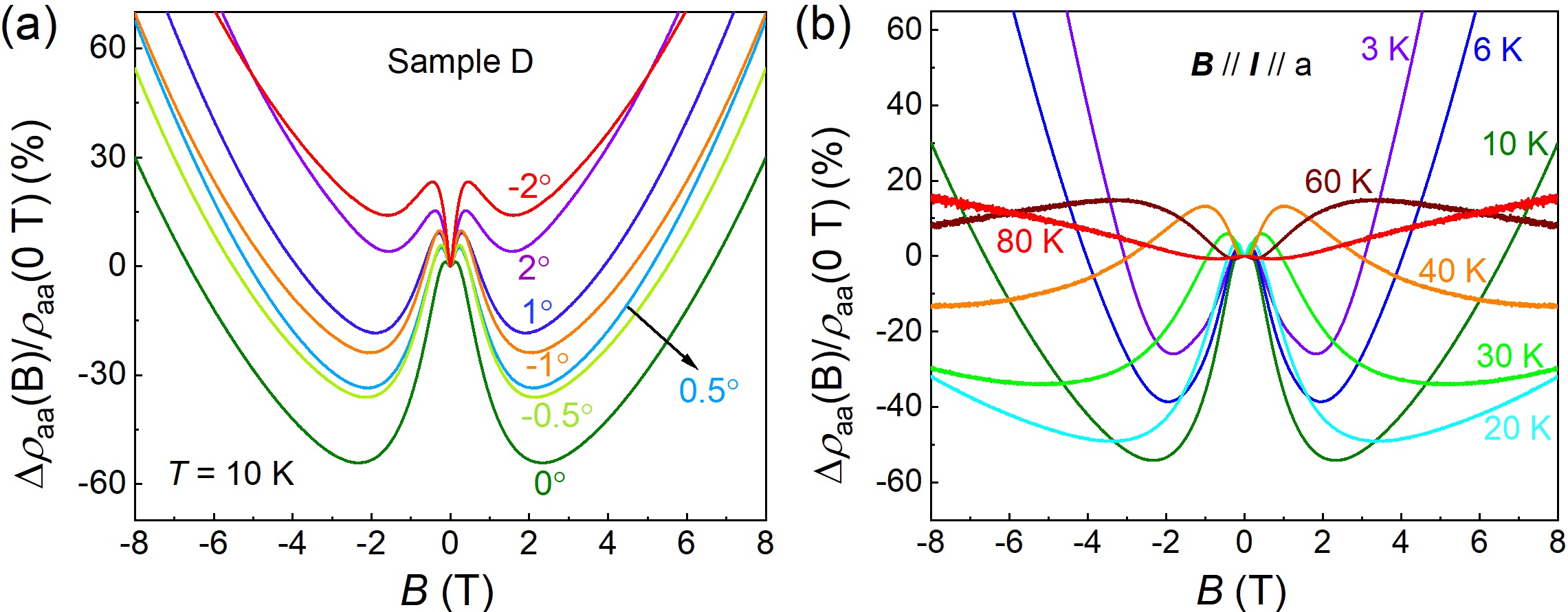}
	\caption{{\bf LMR in sample D.} 
	(a) Magnetoresistance in $\rho_{aa}$ measured at 10 K in sample D for varying $\theta$ near the $a$-axis. 
	(b) LMR for $\boldsymbol{B} \parallel \boldsymbol{I} \parallel a$-axis ($\theta$ = 0$^\circ$) measured at various temperatures.}
\end{figure}

\subsection{Reproducibility of the PHE in sample D}

The unconventional PHE was also reproduced in sample D as shown in Fig. S3. Here, the amplitude of the PHE becomes maximum at $\sim 10$ K [Fig. S3(a)], which seems to be correlated with the temperature where the negative MR becomes most prominent. 
The nearly-regular $\cos \theta$ dependence in $\rho_{ba}^{\rm antisym}$ observed in sample A is also reproduced in sample D [Fig. S3(c)].
Intriguingly, the sign of the $\rho_{ba}^{\rm antisym}$ component in sample D is opposite to that in sample A.

\begin{figure}[h]
	\centering
	\includegraphics[width=12cm]{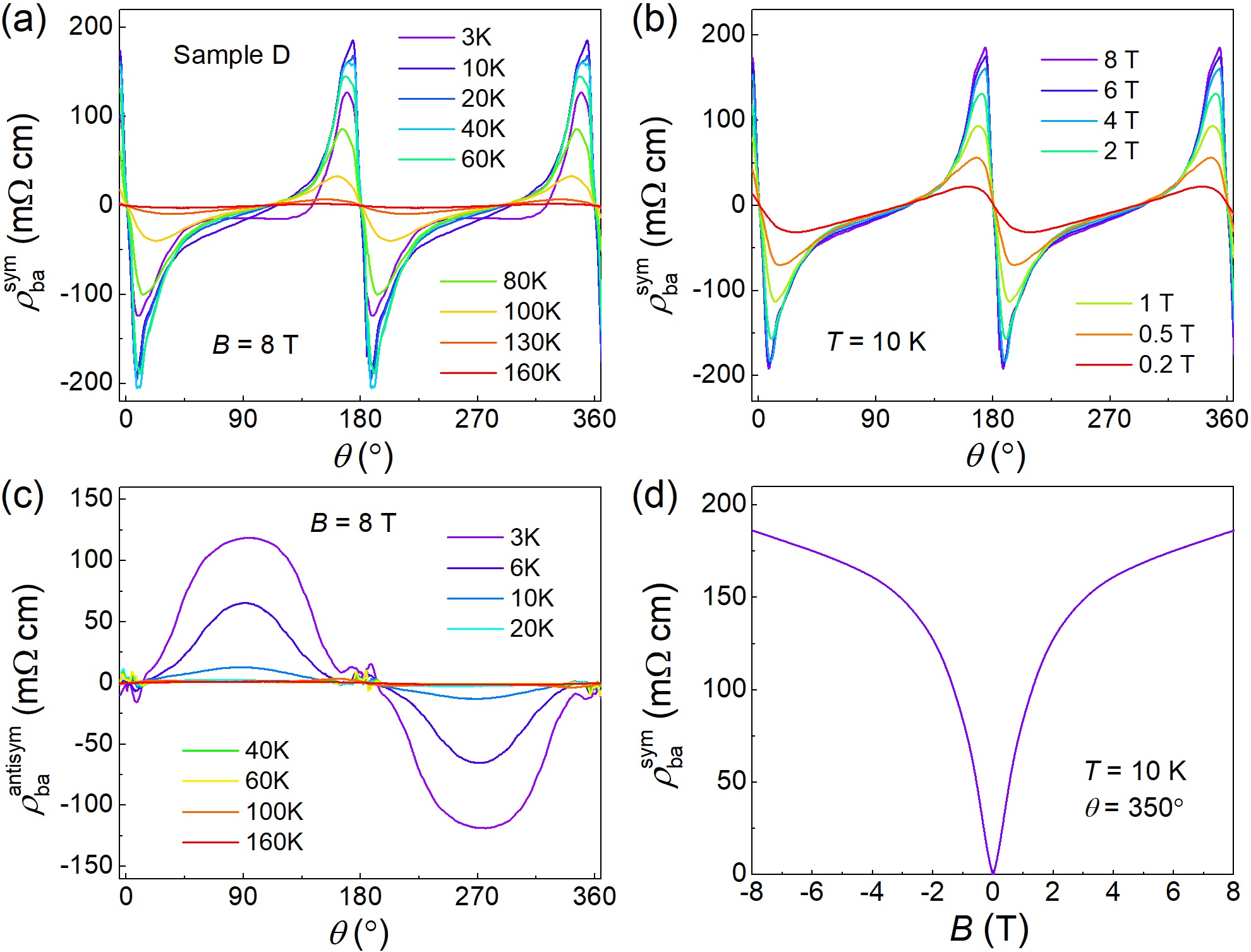}
	\caption{{\bf PHE in sample D.} 
	(a) $\theta$-dependence of $\rho_{ba}^{\rm sym}$ in sample D measured in 8 T at various temperatures. 
	(b) $\theta$-dependence of $\rho_{ba}^{\rm sym}$ at 10 K for varying strengths of $B$. 
	(c) $\theta$-dependence of $\rho_{ba}^{\rm antisym}$ in 8 T at various temperatures. 
	(d) $B$-dependence of the $\rho_{ba}^{\rm sym}$ at 10 K for $\theta = 350^\circ$, which is near the peak position of the $\theta$-dependence of $\rho_{ba}^{\rm sym}$.}
	\label{fig:S3}
\end{figure}

\newpage
\subsection{PHE in a high $T_p$ sample}

We also investigated the PHE in sample E which had $T_p \approx$ 138 K [Fig. S4(a)]. It was found that the PHE still shows an unconventional $\theta$-dependence, but the peak-and-dip structure is not as sharp as that of the $T_p$ = 0 K samples [see Fig. S4(b)]. This is possibly due to a weaker anisotropy in the Fermi velocity, which needs to be confirmed in future studies.

\begin{figure}[h]
	\centering
	\includegraphics[width=11.5cm]{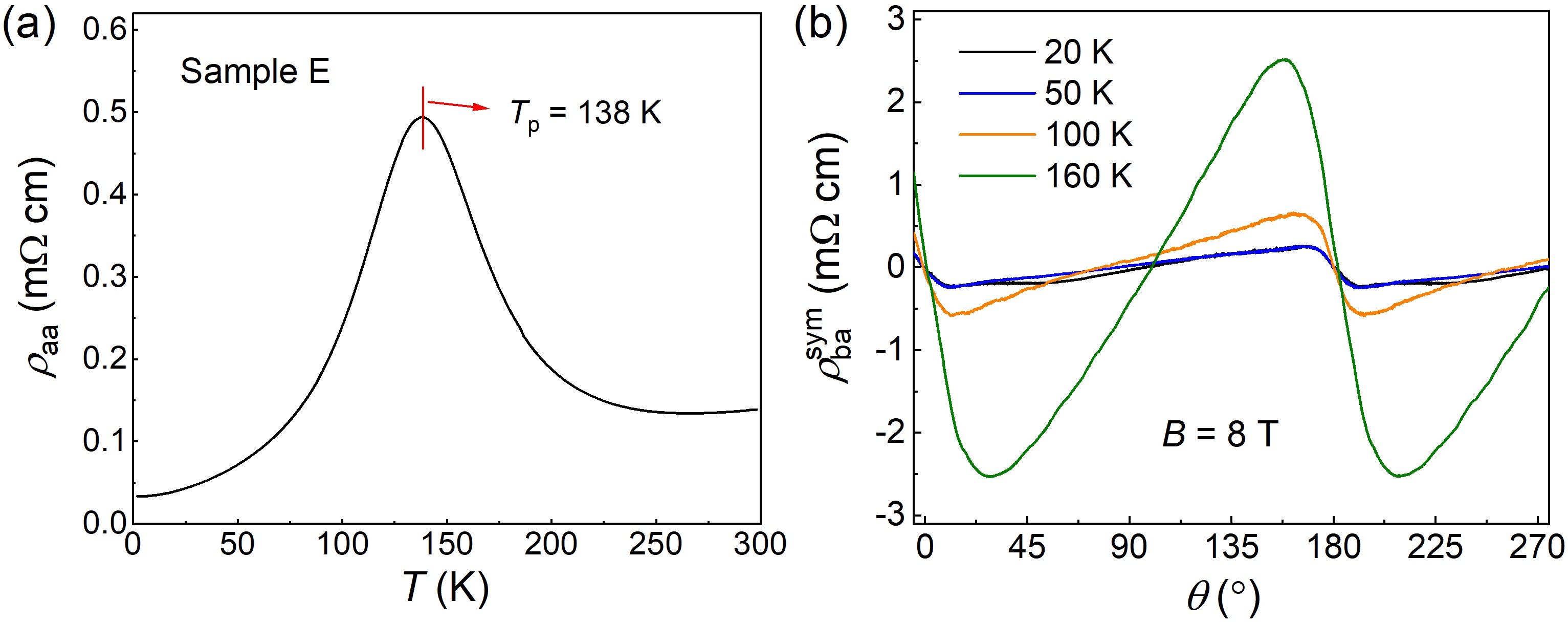}
	\caption{{\bf PHE in sample E.} 
	(a) Temperature dependence of $\rho_{aa}$ in sample E, which shows $T_p$ of 138 K. 
	(b) $\theta$-dependence of $\rho_{ba}^{\rm sym}$ in 8 T at various temperatures.}
	\label{fig:S4}
\end{figure}

\subsection{Measurements of the $b$-axis resistivity $\rho_{bb}$}
\label{sec:fluxTF}
\begin{figure}[h]
	\centering
	\includegraphics[width=12cm]{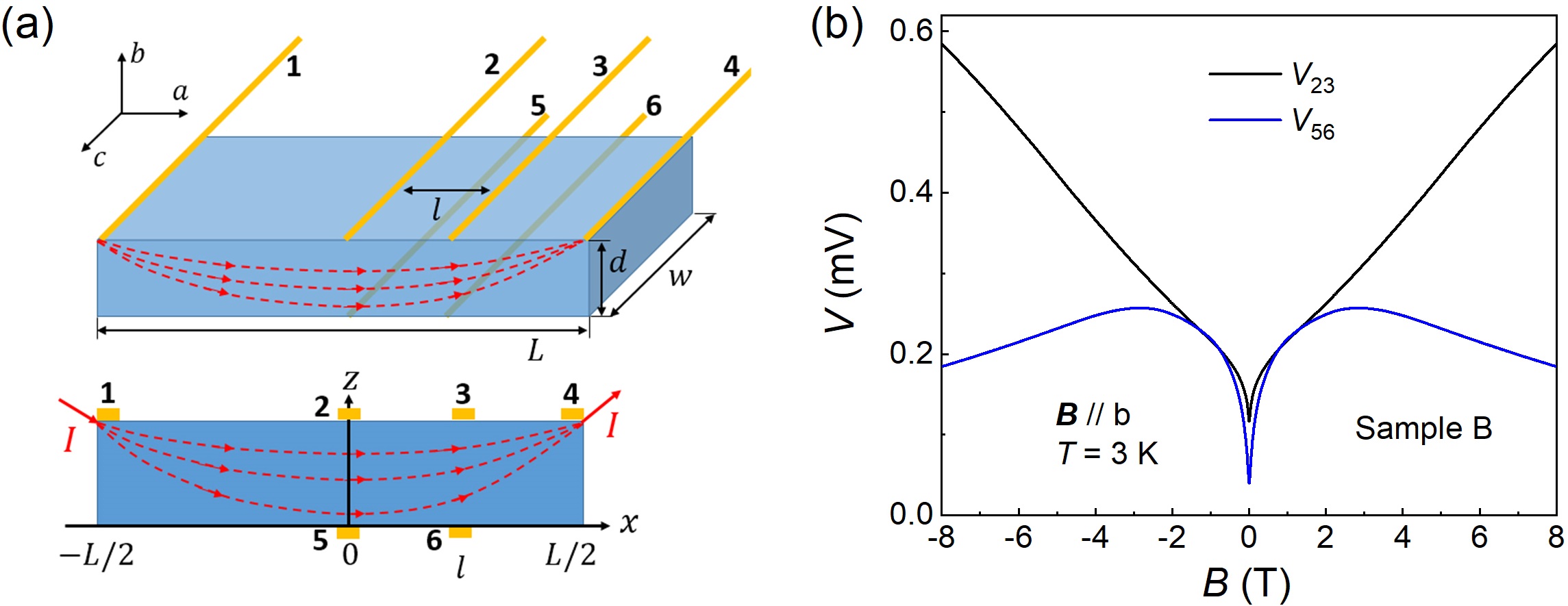}
	\caption{(a) Schematic of the measurement configuration of the flux-transformer method. (b) $B$-dependencies of the voltages $V_{23}$ and $V_{56}$ measured in sample B for $\boldsymbol{B} \parallel b$-axis at 3 K.}
	\label{fig:S5}
\end{figure}

The anisotropic resistivities were measured with the flux-transformer method \cite{Busch1992, Levin1997}, which is illustrated in a schematic in Fig. S5(a). A total of six contact electrodes are made on the top and bottom surfaces, and the voltages appearing at the contact pairs 2-3 and 5-6 are analyzed using a suitable formula considering the geometry \cite{Levin1997}.

We prepared sample B with length of $L$ = 850 $\mu$m, width of $w$ = 152 $\mu$m, and thickness of $d$ = 70 $\mu$m for these measurements. The lower panel of Fig. S5(a) shows a side view of the measurement configuration, where a coordinate system $xz$ is defined on the $ab$ plane. The current contacts 1 and 4 are located at $(-L/2, d)$ and $(L/2, d)$, respectively. The voltage contacts 2, 3, 5, 6 are at $(0, d)$, $(l, d)$, $(0, 0)$, $(l, 0)$, respectively, where $l$ = 263 $\mu$m. The current distribution in the $ab$ plane is depicted in Fig. S5(a), which is determined by the Laplace’s equation \cite{Levin1997}. 

In our analysis, the resistivity anisotropy ratio $\alpha \equiv (\rho_{bb}/\rho_{aa})^{1/2}$ is approximated as $\alpha \approx v_{{\rm F},a}/v_{{\rm F},b} \approx 15.9$ \cite{Wang2022}. 
This leads to $\exp(\pi d \alpha /L) \approx$ 62.8.
The voltages $V_{23}$ and $V_{56}$ in the configuration shown in Fig. S5(a) are given by \cite{Levin1997}:
\begin{equation} 
\begin{split} 
V_{23} \approx \frac{2 I(\rho_{aa} \rho_{bb})^{1 / 2}}{\pi w} \ln \tan \left(\frac{\pi}{4}+\frac{\pi l}{2 L}\right) \label{eq1}
\end{split}
\end{equation}
\begin{equation} 
\begin{split} 
V_{56} \approx \frac{4 I(\rho_{aa} \rho_{bb})^{1 / 2}}{\pi w}\frac{\sin (\pi l / L)}{\sinh [\pi d \alpha / L]} . 
\label{eq2}
\end{split}
\end{equation}
We calculate $\rho_{bb}$ and $\rho_{aa}$ by solving these equations using the data of $V_{23}$ and $V_{56}$ shown in Fig. S5(b), and the results are shown in Fig. 5(a) of the main text.

\begin{figure}[h]
	\centering
	\includegraphics[width=13cm]{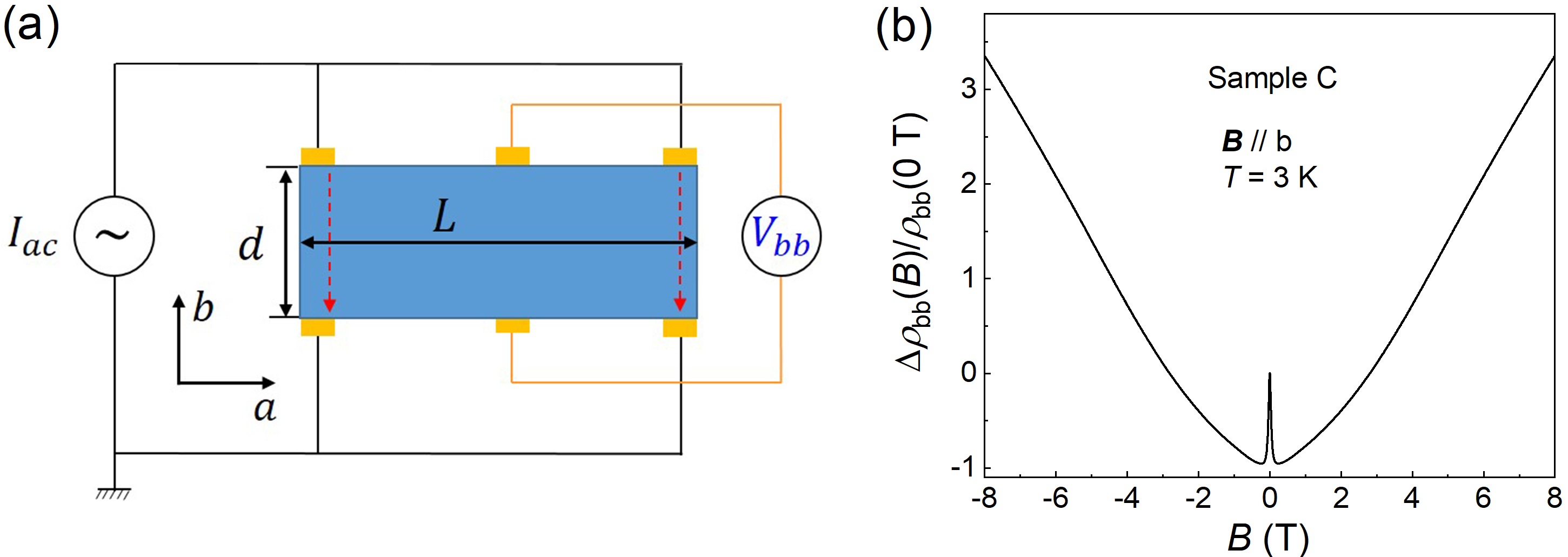}
	\caption{(a) Measurement configuration of sample C, whose thickness $d$ and length $L$ were about 70 $\mu$m and 160 $\mu$m, respectively. (b) Magnetoresistance in $\rho_{bb}$ in sample C for $\boldsymbol{B} \parallel b$-axis at 3 K.  }
	\label{fig:S6}
\end{figure}

To check for the correctness of the magnetoresistance in $\rho_{bb}(B)$ for $\boldsymbol{B} \parallel b$-axis obtained from the flux-transformer method, we performed a direct measurement of the magnetoresistance in $\rho_{bb}$ on another sample with $T_p$ = 0 K (sample C). A schematic of the measurement configuration for sample C is shown in Fig. S6(a). A total of six contact electrodes were made on the top and bottom surfaces of the sample, and they were wired in the way shown in the figure. The contact pair in the middle was used to detect the voltage drop between the top and bottom surfaces, $\Delta V$, and an apparent $\rho_{bb}$ was calculated from $\Delta V / I_{ac}$. 
Due to the decay in the current distribution in the horizontal direction, this method generally gives an apparent resistivity value that is smaller than the true value, but the normalized magnetoresistance is reliable. The result is shown in Fig. S6(b), where the normalized magnetoresistance, namely, the ratio of $\Delta \rho_{bb}(B)/\rho_{bb}(0\,\text{T})$ which cancels the uncertainty in the absolute magnitude, is plotted. This magnetoresistance is consistent with the result obtained from the flux-transformer method.

\section{Landau levels and self-consistent Born approximation}\label{app:theory}

\subsection{Model}\label{app:th:model}

The lowest-energy bands of the nodal line semimetal are described by the $4 \times 4$ Hamiltonian \cite{Wang2022,Wang2023}
\begin{align}\label{eq:th:model}
    H= \begin{pmatrix}
    0 & -v_c \hbar \partial_c -iv_a \hbar \partial_a + \Delta & 0 & - i v_b \hbar \partial_b \\
    v_c \hbar \partial_c - iv_a \hbar \partial_a + \Delta & 0 & - i v_b \hbar \partial_b & 0 \\
    0 & -i v_b \hbar \partial_b &  0  & -v_c \hbar \partial_c + iv_a \hbar \partial_a + \Delta \\
    -i v_b \hbar \partial_b & 0 & v_c \hbar \partial_c + iv_a \hbar \partial_a + \Delta & 0
    \end{pmatrix},
\end{align}
where we use the parameters $v_a = 6.6 \times 10^5$ ms$^{-1}$, $v_b = 0.45 \times 10^5$ ms$^{-1}$, $v_c = 1.8\times 10^5$ ms$^{-1}$, $\Delta = 0.018$\,eV as in Ref.~\cite{Wang2022}.

Next, we consider the effect of a magnetic field. The following formulas are given for a field in the $b$ direction. Thus, $k_\|=k_b$ is a good quantum number and  the Hamiltonian can be written in terms of ladder operators \cite{Wang2023}
\begin{align}\label{Hb}
    H_{k_b}= \begin{pmatrix}
    -\frac{1}{2} g \mu_B B & \Omega_c \hat a + \Delta & 0 & v_b  \hbar k_b \\ 
    \Omega_c \hat a^\dagger + \Delta & -\frac{1}{2} g \mu_B B & v_b\hbar k_b & 0 \\
    0 &  v_b\hbar k_b  & \frac{1}{2} g \mu_B B & -\Omega_c \hat a^\dagger + \Delta \\
    v_b\hbar k_b & 0 & - \Omega_c \hat a + \Delta & \frac{1}{2} g \mu_B B
    \end{pmatrix},
\end{align}
where the g-factor is $g = 20$ ($g = 4$) for magnetic fields along the $b$-axis ($a$-axis).
Introducing the quantum number $m$ with $a^\dagger a|m\rangle=m|m\rangle$, $m=0,\dots,m_\text{max}$, allows to write the Hamiltonian and the Green function
\begin{align}
    G_{k_b}(\omega)=\left(\hbar\omega - H_{k_b}-\Sigma\right)^{-1},
\end{align} as a $4 m_\text{max} \times 4 m_\text{max}$ matrix with indices $n,m$ and $n',m'$, which we compute numerically. Note that each Landau level (LL) carries an extra degeneracy $N_\text{LL}$, which counts the number of flux quanta through the system.
The area per flux quantum is $2 \pi l_B^2$ where $l_B=\sqrt{\hbar/(e B)}$ is the magnetic length, thus $N_\text{LL}=A/(2 \pi l_B^2)$, where $A$ is the area perpendicular to $B$.

The local disorder potential $V_n(\vec r)$ within band $n$ is modeled by a random potential with
\begin{align}
    \langle V_n(\vec r) V_{n'}(\vec r')\rangle=\gamma \delta(\vec r-\vec r')\delta_{n,n'}.
\end{align}
For all plots we use  $\gamma \approx 0.06\, \text{nm}^3 (\text{eV})^2$, roughly reproducing the experimentally observed resistivities. Within self-consistent Born approximation the self-energy is given by
\begin{align}
    \Sigma_{n,m,n',m'}(\omega)=\delta_{m,m'} \frac{\gamma}{2 \pi l_B^2} \,\int \frac{dk_b}{2 \pi} \sum_{\tilde m} G^{n,\tilde m,n',\tilde m}_{k_b}(\omega),
\end{align}
where the factor $1/(2 \pi l_B^2)$ arises from the LL degeneracy discussed above.
Note that the self-energy acts as an identity matrix in the LL indices $m$ and $m'$ but not in the  band indices $n$ and $n'$. 

The longitudinal dc conductivity is calculated from the Kubo formula at zero temperature.
\begin{align}\label{kuboFull}
    \sigma_{bb}=\frac{\hbar e^2}{2 \pi l_B^2} \frac{1}{\pi} \,\int \frac{dk_b}{2 \pi} \text{tr}\!\left[\text{Im}\, G_{k_B}(\omega=\mu)\, \hat v \,\text{Im}\, G_{k_B}(\omega=\mu) \, \hat v\right] ,
\end{align}
where $\hat v=\frac{\partial H_{k_B}}{\hbar \partial k_b}$ is the velocity operator.
Note that the formula neglects vertex corrections, discussed separately in Sec.~\ref{sec:vertex}.
Importantly, $\sigma_{bb}$ strongly depends on the value of the chemical potential, which changes a function of the magnetic field.
The chemical potential is determined from the condition
\begin{align}
 n= \frac{\hbar}{2 \pi l_B^2}  \,\int \frac{dk_b}{2 \pi} \int_{0}^\mu \frac{d\omega}{\pi}\text{tr}\!\left[\text{Im}\, G_{k_B}(\omega)\right] ,
\end{align}
where $n$ is the electron density measured relatively to the charge neutrality point ($\mu=0$), where we use that the model is particle-hole symmetric.

As discussed in the main text, we have to also take into account the effects of large-scale inhomogeneities (charge puddles) arising mainly from charged impurities in a system with an extremely low electron density.
We model this effect phenomenologically assuming that the electron density is not fixed but instead obtains a Gaussian distribution, 
\begin{align}p(n)\sim e^{-\frac{1}{2} (n-\bar n)^2/\sigma_n^2},\label{eq:averaging}
\end{align}
where for our plots we use $\bar n=1.1 \times 10^{16} \text{ cm}^{-3}$, $\sigma_n \approx 0.3 \times 10^{16} \text{ cm}^{-3}$. 

To average over $n$, we use that the resistivies are highly anisotropic, $\rho_{bb} \gg \rho_{aa}, \rho_{cc}$. Thus, transport in the $b$ direction (in the $a$ direction) is effectively governed by a network of resistors connected in parallel (in series), respectively. Thus, we approximate
\begin{align}\rho_{aa} \approx \int \rho_{aa}(n) p(n) dn, \qquad \rho_{bb} \approx \left(\int \sigma_{bb}(n) p(n) dn\right)^{-1}.
\end{align}

\subsection{Longitudinal conductivity in the chiral-anomaly regime of ZrTe$_5$}

Next, we develop an analytical theory of transport in the regime where the longitudinal resistivity shows a pronounced dip. 

Upon increasing $B$, LLs form and the chemical potential $\mu$ becomes smaller and smaller. The dip occurs always in the regime, when $\mu$ crosses the dispersion of the lowest LL only. At the same time, we demand that the chemical potential has not yet reached the flat band regime. 
The parameter regime is marked by the shaded region in Fig. 1(b) and Fig. 4 of the main text. This field range corresponds to the chiral-anomaly regime discussed in the main text.

As the scattering rate $\Gamma \simeq$ 0.2 meV, see Fig. 4(a) of the main text, is at least an order of magnitude smaller than the distance of LLs in this regime and the calculations in Sec. IB of the main text apply. 

As only the LLL contributes, one can describe the Green function and scattering rates are given by Eq. (4) and (5) of the main text, respectively, where the factors $\alpha^\pm$ are geometric factors arising from the projection to the two Fermi points of the lowest Landau level. They are given by
\begin{align} 
\alpha^\pm&= \sum_{n,n',m,m'}\langle \pm k_F  |n,m'\rangle\langle n,m| k_F\rangle \langle k_F |n', m\rangle \langle n',m' | \pm k_F\rangle,
\end{align}
where $\alpha^{+}$ ($\alpha^-$) encodes the overlap for momenta with the same (opposite) Fermi momentum. Here $|n,m\rangle$ are the Landau-level eigenstates discussed above, which  are computed by diagonalizing $H$, Eq.~\eqref{Hb}.
For $B$ = 0.2 T in the $b$ direction, we obtain, for example, 
\begin{align}\label{eq:parametersCA}
    \alpha^+\approx 0.78, \quad \alpha^-\approx 0.05.
\end{align}
Taking $v_F$ from the range depicted in Fig. 4(b), one can compare the analytical result
\begin{align}
    \Gamma_{k_F}&\approx \Gamma^+_{k_F}+\Gamma^-_{k_F}=\frac{\gamma}{v_F \hbar 4 \pi l_B^2} (\alpha^{+}+\alpha^{-}) \label{eq:gammaTot}
\end{align}
and
\begin{align}
    \sigma_{bb}\approx 2 \frac{e^2 \hbar v_F^2}{2 \pi l_B^2} \frac{1}{\pi} \int \frac{dk}{2 \pi} \left(\text{Im}\, g_k(\omega=0)\right)^2\approx \frac{e^2 v_F}{4 \pi^2 l_B^2 \Gamma_{k_F}}\label{sigmabb}
\end{align}
to the full evaluation of the Kubo-formula, Eq.~\eqref{kuboFull}, in self-consistent Born approximation. The analytical formula ~\eqref{sigmabb} predicts
$0.11 \, \Omega \text{cm} \leq \rho_{bb} \leq 0.19\,\Omega \text{cm}$
for the longitudinal resistivity in the chiral anomaly regime, which is in quantitative agreement with the numerical results, as indicated by the gray-shaded region in Fig. 4(d).
%we find in this regime the expected quantitative agreement with an error of less than $10\%$.
In passing, using Eq.~\eqref{eq:gammaTot}, we can rewrite Eq.~\eqref{sigmabb} in a form where all $l_B$ factors cancel
\begin{align} \label{condFinal_noVertex}
    \sigma_{bb}\approx \frac{e^2}{2 \pi \hbar} \frac{v_F^2 \hbar^2}{ \gamma (\alpha^+ +\alpha^-)/2}.
\end{align}

\subsection{Validity of the theory and role of vertex corrections} \label{sec:vertex}

While our theory reproduces the emergence of negative magnetoresistance and the associated characteristic dips in the longitudinal resistivity as function of fields, there are both quantitative and qualitative disagreements between theory and experiment. In the following, we provide a concise discussion of the validity of the theoretical approach and why it is expected to break down in certain regimes.

Our theory treats short-range impurities within self-consistent Born approximation, while large-scale inhomogeneities (electron-hole puddles) arising from charged defects are only taken into account by averaging the final results over densities, see Sec.~\ref{app:th:model}.
A clear experimental indication for the importance of puddles is that the scattering rate obtained from transport is almost an order of magnitude larger than obtained from quantum oscillation measurements
\cite{Wang2022}. Any fully quantitative description of transport has therefore to include a theory of puddle formation governed by Coulomb interactions and of the impact of large scale inhomogeneities on magnetotransport, which is a very challenging problem beyond the scope of our study.
While averaging over densities is nominally justified when the electronic mean free path is much smaller than the size of puddles, this is most likely not the case in our system and thus we do not expect a full quantitative agreement of theory and experiment. 

Another shortcoming of our theory is that we are restricted to magnetic fields larger than $\sim$ 0.1 T as more and more Landau levels contribute for smaller fields. Therefore, it is not clear, whether it can reproduce the negative magnetoresistance observed for very small fields for $\boldsymbol{B} \parallel b$ (but not for $\boldsymbol{B} \parallel a$). In Fig. 4(c) of the main text, a small maximum is visible for the smallest fields, which is not observed in the experiment. This feature does, however, depend very sensitively on the density distribution $p(n)$ used for averaging.

An advantage of the self-consistent Born approximation is that it includes the effect of spectral broadening and thus gives `reasonable' results for a wide set of regimes, e.g., when bands become flat. It does, however, not include interference effects responsible for Anderson localization, which likely become important when the impurity scattering rate is much larger than the relevant bandwidth of the system. In our system this occurs for very large magnetic field, where our theory predicts a drop of the longitudinal resistivity, which is not observed experimentally. This discrepancy is, perhaps, explained by the onset of localization physics in the experimental system.

Finally, when computing transport using Eq.~\eqref{kuboFull}, vertex corrections have been neglected to simplify the numerics. A well-known artifact of such a calculation is that in this case the transport-relaxation rate (describing the relaxation of current) is replaced by the single-particle relaxation rate.
In chiral-anomaly region, where only a single channel contributes to transport and where self-consistency effects can be neglected, we can easily compute the correct transport relaxation rate. 
The single-particle scattering rate counts the rate with which particles are scattered away, without taking into account their final state. Here, the rate $\Gamma^+_{k_F}$ in Eq.~\eqref{eq:gammaTot} describes scattering from, e.g., the Fermi momentum $\pm k_F$ to the same momentum $\pm k_F$ (or, more precisely, to momenta close by). Such scattering events do not change the velocity of the electron and therefore $\Gamma^+$ processes do not contribute to the current relaxation. In contrast, the rate $\Gamma_{k_F}^-$ describes the scattering from $k_F$ to $-k_F$, which changes the velocity by $2 v_F$, and thus contributes to current relaxation with a factor $2$. Taking this into account, Eq.~\eqref{condFinal_noVertex}, has to be replaced by 
\begin{align} \label{condFinal_withVertex}
    \sigma_{bb}\approx \frac{e^2}{2 \pi \hbar} \frac{v_F^2 \hbar^2}{ \gamma \alpha^-}.
\end{align}
Using the parameters shown in Eq.~\eqref{eq:parametersCA}, this implies that the conductivity in the chiral-anomaly regime is in this case a factor of $8$ bigger than predicted by our numerics. This is an surprisingly large effect reflecting a substantial suppression of backscattering within our model.

Averaging over densities as described by Eq.~\eqref{eq:averaging} will reduce the effect, but we conclude that our calculation substantially underestimates the  conductivity in the chiral-anomaly region. Thus, it underestimates the negative magnetoresistance in this parameter range.

\section{Simulation of the planar Hall effect from the data of $\rho_{aa}(B)$ and $\rho_{bb}(B)$}

%Understood:\AR{WE CHANGED xy TO ab BOTH HERE AND IN THE MAIN TEXT. I ALSO PUT BACK THE ARGUMENT FOR THE RESCALING OF $B$ USING THE VECTOR POTENTIAL AS I DID NOT SUCCEED TO MAKE YOUR ARGUMENT USING MAXWELL'S EQUATION PRECISE. I ALSO  ADDED A SUBSTANTIAL DISCUSSION ON THE VALIDITY OF THE RESCALING ARGUMENT}

As discussed in the main text, we rescale the coordinates and momentum in ZrTe$_5$ to make the rescaled Fermi surface isotropic in the $ab$ (i.e., $xz$) plane. This can be done with the rescaling $r_a' = r_a$, $r_b' = \lambda r_b$ and $k_a' = k_a$, $k_b' = k_b / \lambda$ with the scaling factor $\lambda \equiv v_{a}/v_{b}$ such that $v_b k_b=v_a k_b'$. 
With the rescaling matrix 
\begin{equation} 
\boldsymbol{\lambda} \equiv
\left[\begin{array}{cc}
1 & 0 \\
0 & \lambda
\end{array}\right]\, , \label{eq2}
\end{equation}
the current density $\boldsymbol{j}$ is rescaled to $\boldsymbol{j'} = \boldsymbol{\lambda} \boldsymbol{j}$  because $\boldsymbol{j}=-ne\dot{\boldsymbol{r}}$. On the other hand, the electric field $\boldsymbol{E}$ is rescaled to $\boldsymbol{E'} = \boldsymbol{\lambda}^{-1} \boldsymbol{E}$ because $E_{i}=-\frac{\partial V_{i}}{\partial r_{i}}$ where $V_i$ is the voltage. Therefore, the Ohm’s law in the rescaled system, $\boldsymbol{\rho'} \boldsymbol{j'} = \boldsymbol{E'}$, is written as
\begin{equation} 
\boldsymbol{\rho}^{\prime} \boldsymbol{\lambda} \boldsymbol{j}
= \boldsymbol{\lambda}^{-1} \boldsymbol{E}\, ,
\end{equation} 
from which we conclude $\boldsymbol{\rho'} = \boldsymbol{\lambda^{-1}}\boldsymbol{\rho}\boldsymbol{\lambda^{-1}}$. This means
\begin{equation} 
\left[\begin{array}{cc}
\rho_{a a}' & \rho_{a b}' \\
\rho_{b a}' & \rho_{b b}'
\end{array}\right]=\left[\begin{array}{cc}
\rho_{a a} & \rho_{a b} / \lambda \\
\rho_{b a} / \lambda & \rho_{b b} / \lambda^{2}
\end{array}\right]. \label{eq1}
\end{equation}

The rescaling of a magnetic field in the $ab$ plane is not very intuitive. Using a gauge where $\vec A= r_c\vec B \times \hat{\vec c} $ and using that $\vec A$ enters Eq.~\eqref{eq:th:model} by minimal substitution, $-i \hbar \partial_i \to -i \hbar \partial_i-e A_i $, it follows that $A_b'=A_b/\lambda$, $A_a'=A_a$ and thus $B_a' = 
B_a / \lambda$, $B_b' = B_b$. 
This rescaling does not apply to the Zeeman terms, which give, however, only subleading contributions.

The rescaling also affects the disorder potentials. If disorder potentials are short-ranged compared to the very long electronic wavelength in our system, one can approximate them by $\delta$-potentials, which leads to isotropic scattering in the rescaled coordinate system. Note that this argument does not necessarily apply to larger-scale inhomogeneities (electron-hole puddles) created from the screening of charged impurities. Experimentally, we find, however, that the rescaling works remarkably well to explain the pronounced resistivity anisotropies in our system. This  suggests that main features of transport can be understood from short-range scattering as assumed in our theoretical analysis.

\vspace{3mm}
For $\boldsymbol{B}$ at angle $\theta$  in the $ab$-plane, the rescaled magnetic field is $\boldsymbol{B}'=B[\cos (\theta) / \lambda, \sin (\theta)]$.
% with the amplitude of $B'=B \sqrt{\cos ^{2} \theta / \lambda^{2}+\sin ^{2} \theta}$. 
The effective magnetic-field angle $\theta'$ in the projected isotropic space is calculated with $\tan (\theta')=\lambda \cdot \tan (\theta)$, which implies that there is a non-linear relationship between $\theta$ and $\theta'$. 
%The rescaled magnetic field and rotation angle can explain the unusual angular dependence of the PHE in ZrTe$_5$, which can be clearly seen with rescaling resistivity.
%In quantum mechanics, the momentum $\boldsymbol{k}$ changes to $\boldsymbol{k} - e\boldsymbol{A}/\hbar$ in the presence of a magnetic field $\boldsymbol{B}$ derived from a vector potential $\boldsymbol{A}$. This means that when $\boldsymbol{k}$ is scaled by $\lambda$, the magnetic field $\boldsymbol{B}$ must also be scaled by $\lambda$. Hence, the applied magnetic field in the $xz$ plane is rescaled as $\boldsymbol{B}'=B[\cos (\theta) / \lambda, \sin (\theta)]$.
To simulate the $\theta$-dependence of the resistivity tensor, we make a crude assumption that in the projected isotropic system, the way the magnetic-field-induced resistivity anisotropy develops is independent of the orientation of $\boldsymbol {B'}$. 
This means that  the $B'$-dependencies of the resistivities along $\boldsymbol {B'}$ and perpendicular to $\boldsymbol {B'}$ do not change upon rotating $\boldsymbol {B'}$. With this assumption, we can use the experimentally-obtained $\rho_{aa}(B)$ and $\rho_{bb}(B)$ for $\boldsymbol {B} \parallel b$-axis ($\theta' = \theta$ = 90$^\circ$) as the input of the simulation, and calculate the rotated resistivity tensor by using the rotation matrix $\boldsymbol {R}$ which is given by 
\begin{equation} 
\boldsymbol {R}(\alpha)=\left[\begin{array}{cc}
\cos (\alpha) & -\sin (\alpha) \\
\sin (\alpha) & \cos (\alpha)
\end{array}\right] \label{eq4}
\end{equation}
for a rotation angle $\alpha$. 
%Note that when $\boldsymbol {B}$ is parallel to $a$- or $b$-axis, the original resistivity-anisotropy axes remain unrotated. 
We chose $\boldsymbol {B} \parallel b$-axis for the reference measurement of $\rho_{aa}(B)$ and $\rho_{bb}(B)$, because the resistivity is stable with a slight tilt of $\boldsymbol {B}$ from $b$ [see Fig. 3(a) of the main text]; on the other hand, the resistivity changes a lot with a slight tilt of $\boldsymbol {B}$ from $a$, so $\boldsymbol {B} \parallel a$-axis is less suitable as a reference.

The projected resistivity tensor $\boldsymbol{\rho'}$ for this $\boldsymbol {B} \parallel b$-axis situation (for which $\theta' = \theta$ = 90$^\circ$) is
\begin{equation} 
\boldsymbol{\rho'}(\theta'=90^\circ, B')=\left[\begin{array}{cc}
\rho_{aa}(B') & 0 \\
0 & \rho_{bb}(B') / \lambda^{2}
\end{array}\right]. \label{eq3}
\end{equation}
Since this $\boldsymbol{\rho'}$ takes $\theta' = 90^\circ$ as the angle of reference, the rotation angle $\alpha$ of the magnetic field should be measured from $\theta' = 90^\circ$, which means that $\boldsymbol{\rho'}(\theta')$ for an arbitrary angle $\theta'$ is obtained by rotating Eq.~\eqref{eq3} with $\boldsymbol {R}(\theta'- 90^\circ)$. Hence, we obtain
\begin{equation}
\boldsymbol{\rho'}(\theta', B')=\boldsymbol {R}(\theta'-90^{\circ}) \boldsymbol{\rho'}(\theta'=90^{\circ}, B') \boldsymbol {R}^{-1}(\theta'-90^{\circ})\, .
\end{equation}
The resistivity tensor $\boldsymbol{\rho}(\theta, B)$ in the original unscaled system can be obtained by back-scaling $\boldsymbol{\rho'}$ via 
$\boldsymbol{\rho}(\theta, B)= \boldsymbol{\lambda} \boldsymbol{\rho'}(\theta', B') \boldsymbol{\lambda}$, and the result is
\small
\begin{equation}
\boldsymbol{\rho}(\theta, B)
= \left[\begin{array}{ll}
\Delta \rho \cdot \cos ^{2} \theta'+\rho_{a a}(B') & \lambda \Delta \rho \cdot \sin \theta' \cos \theta' \\
\lambda \Delta \rho \cdot \sin \theta' \cos \theta' & \lambda^2 \Delta \rho \cdot \sin ^{2} \theta' + \lambda^2 \rho_{a a}(B')
\end{array}\right], \label{eq5}
\end{equation}
\normalsize
where $\theta'= \tan^{-1} \left( \lambda \cdot \tan (\theta) \right)$, $\Delta \rho = \rho_{bb}(B') / \lambda^2 - \rho_{aa}(B')$, and $B'=B \sqrt{\cos ^{2} \theta / \lambda^{2}+\sin ^{2} \theta}$.

\bibliography{ZrTe5_bibliography}